**Strategies for the Simulation of Sea Ice Organic Chemistry: Arctic Tests and Development**

S. Elliott[1], N. Jeffery[1], E. Hunke[1], C. Deal[2], M. Jin[2], S. Wang[1], E. Elliott Smith[3], S. Oestreicher[4]

[1]Climate Ocean Sea Ice Modeling (COSIM), Los Alamos National Laboratory, Los Alamos, NM 87545

[2]International Arctic Research Center, Fairbanks, AK 99775

[3]Biology Department, University of New Mexico, Albuquerque, NM 87131

[4]Applied Mathematics, University of Minnesota, Minneapolis, MN 55455

**Abstract**. A mechanism connecting ice algal ecodynamics with the buildup of organic macromolecules in brine channels is tested offline in a reduced model of pack geochemistry. Driver physical quantities are extracted from the global sea ice dynamics code CICE, including snow height, column thickness and internal temperature. The variables are averaged at the regional scale over ten Arctic biogeographic zones and treated as input matrices at four vertical habitat levels. Nutrient-light-salt limited ice algal growth is computed along with the associated grazing plus mortality. Vertical transport is diffusive but responds to pore structure. Simulated bottom layer chlorophyll maxima are reasonable, though delayed by about a month relative to observations. This highlights major uncertainties deriving from snow thickness variability. Upper level biota are generated intermittently through flooding. Macromolecular injections are represented by the compound classes humics, proteins, polysaccharides and lipids. The fresh biopolymers behave in a successional manner and are removed by bacteria. In baseline runs, organics are introduced solely through cell disruption, and internal carbon is biased low. Continuous exudation is therefore appended, and agreement with dissolved organic or individual biopolymer measurements is achieved when strong release is coupled to light availability. Detrital carbon then reaches hundreds of micromolar, sufficient to support physical changes to the ice matrix. Through this optimized model version we address the question, are high molecular weight organics added to the brine network over and above background spillage? The mechanism is configured for ready extension to the Antarctic, so that global ice organic chemistry issues can be targeted.

**To whom correspondence should be addressed**: Scott Elliott, sme@lanl.gov

**1. Introduction**

Pack ice plays a significant role in establishing the global climate state, acting through a variety of biophysical mechanisms and feedbacks. For example, chlorophyll absorption at the ice domain edge can amplify global warming (Lengaigne et al. 2009) while the loss of reflective coverage dramatically alters planetary albedo (Screen and Simmonds, 2010). Organic chemistry is critical to the physical features of sea ice, and so it is also beginning to be considered dynamically in the context of global change (Janech et al. 2006; Riedel et al. 2008; McNeil et al. 2012; Underwood et al. 2013). Brine drainage mechanisms, thermodynamics, habitat volume and nutrient retention are all determined in part by the macromolecular content of internal channels (Underwood et al. 2010; Krembs et al. 2011). Biopolymers are now recognized as a component of the global marine climate system at multiple levels (Burrows et al. 2014; Carpenter and Nightengale, 2015; Letscher et al. 2015). Yet for the pack ice system, reduced carbon geocycling has not been simulated at large scales (see e.g. Tedesco et al. 2012 for an exception that proves the rule). In the present work we apply an approach to the problem that is pan-Arctic in extent, working from several preexisting ecodynamics models developed in our group (Deal et al. 2011; Elliott et al, 2012). Emphasis is placed on organic compound production and loss within the saline network, but in fact all major determinants are set by ice algal nutrients and traditional carbon cycling (Amon et al. 2001; Thomas and Papadimitriou, 2003; Riedel et al. 2008; Aslam et al. 2012a). Hence our scheme is also relevant for more general pack biogeochemistry calculations, and it may be extensible to both hemispheres.

An offline numerical framework is constructed to provide for testing of the new ice organic mechanism. The format is suitable for experimentation and refinement on lap top computers, but transfer to full global climate models is already underway (Hunke et al. 2015). Light absorption and penetration are computed through the pack ice column, along with diffusive tracer transport and continuity for all major nutrients, the biota supported by them and dissolved detritus released as a set of byproducts. Results are analyzed based on observed distributions and properties of natural macromolecules in environmental ice systems (Raymond et al. 1994; Krembs et al. 2002; Janech et al. 2006; Riedel et al. 2008; McNeil et al. 2012; Aslam et al. 2012a and b; Underwood et al. 2013). In addition, validation is performed against chlorophyll data sets for lower habitat layers and also a selection of Arctic ice organic carbon measurements. We explore implications for ice algal adaptation to extreme low temperature brine survival, and speculate on trajectories for the system under upcoming global change (Serreze et al. 2007; Krembs and Deming, 2008; Krembs et al. 2011; Post et al. 2013). For the moment, our core mechanism bypasses phase state distinctions among macromolecules (Underwood et al. 2010). Steps which might be needed to introduce adsorption, colloids and gelling are therefore mentioned briefly in the discussion section (Chin et al. 1998; Janech et al. 2006; Krembs et al. 2011; Underwood et al. 2013).

It is argued that comparisons between our preliminary computations and the available observational data indicate targeted exudation by the brine algae. Thus we directly address the question, are ice biomacromolecules input from the biota additionally to standard background ecodynamic processes (cell disruption)? Direct and intentional introduction into the brine environment appear sufficient to restructure the pore system, enhance nutrient retention, and maintain stratification/position in an attenuating light field (Lavoie et al. 2005; Krembs and Deming, 2008; Underwood et al. 2010). Our results show that selected fresh bioorganic compounds can be injected into the saline network due to growth alone, independent of standard grazing and mortality. We thus identify and explore numerically an alternate ecological flow of carbon, in simulations conducted at the full polar scale. Macromolecular concentrations attained in our calculations reach levels cited in the literature for the control of pore structures and living spaces (Raymond et al. 1994; Janech et al. 2006; Krembs et al 2011). Although organic emissions are treated here through the set of idealized biological tracer classes protein, polysaccharide and lipid (Parsons et al. 1984; Benner, 2002; Underwood et al. 2010; Elliott et al. 2014; Carpenter and Nightingale, 2015), heterogeneous polymeric combinations may be involved as well –glycoproteins, aminosugars, uronic acids, insoluble polysaccharides and short chain humics have all been identified in the brine (Calace et al. 2001; Riedel et al. 2008;

Underwood et al. 2010; Krembs et al. 2011; Aslam et al. 2012a; Underwood et al. 2013). The saline mix may also include chromophoric groups which are fresh, aromatic and mycosporine-like (Norman et al. 2011). Actual molecular configurations may be tailored and specialized for cold-hardiness by strong, continual evolutionary pressure (Janech et al. 2006; Krembs and Deming, 2008; Aslam et al. 2012a and b).

The text is structured as follows: Driver simulations in the parent model CICE are described (Hunke et al. 2015), along with initial and boundary conditions plus ice algal regulators such as snow and ice thickness. All these quantities and more are averaged over regional subdomains defined in recent biogeographies (Carmack and Wassmann, 2006; Deal et al. 2011). Assumptions of the reduced model strategy therefore involve matching with precipitation, thermodynamic and seawater interfacial constraints taken from the parent ice dynamics -all imposed at the synoptic scale. Our approaches to nutrient, light and salt limited growth, to light attenuation, ice-internal transport, porosity dynamics and organochemical kinetics are described from an historical viewpoint (Lavoie et al. 2005; Jin et al. 2006; Vancoppenolle et al. 2010; Deal et al. 2011; Elliott et al. 2012) with mechanistic information provided in several appendices. The reduced code is effectively a set of layered one dimensional population/chemical dynamics calculations, resolved vertically at the scale of a few centimeters. The method is developed here primarily to preview more complete CICE biogeochemistry calculations. We present a mechanism which can also be applied in the Southern Hemispheric ice domain. For example, an ocean-ice iron cycle is included (Van der Merwe et al. 2009; Lannuzel et al. 2010) which is automatically shut down since trace metals tend to be replete in the Arctic (Aguilar-Islas et al. 2007 and 2008). Data collected for validation purposes include sea ice chlorophyll, total dissolved organics and most importantly, a set of chemical analyses specific to the proteins, polysaccharides and aged refractory carbon (e.g. Amon et al. 2001; Krembs et al. 2002; Arrigo, 2003; Riedel et al. 2008; Underwood et al. 2010). For the core ecology, optimization was focused upon pigments in bottom ice segments because field studies are relatively numerous. Organic release mechanisms were then adjusted for the exudates, and average concentrations rose upward toward measurement data.

Core results from the reduced Arctic organic model are presented as plots of time evolution over a typical year of simulation. The major habitat layers bottom, interior, freeboard and infiltration are highlighted. Upper level biota are less well known in the Arctic than for the Southern Ocean ice domain (Ackley and Sullivan, 1994; Gradinger, 1999; Meiners et al. 2012). Our exercise predicts that the topmost habitats may exist in peripheral ecogeographic zones, but with the caveat that they are tenuous –there is a strong sensitivity to snow-ice thickness ratios which determine pack submersion (Fritsen et al. 1998). In a baseline run, organics are supported solely by traditional cell disruption and with the exception of prescribed humics, pack internal carbon remains low relative to observations (Amon et al. 2001; Calace et al. 2001; Krembs et al. 2002). A conclusion is that exudation from the brine algae could be required, taking place over and above simple release through mortality and grazing. A scheme specifying threshold light levels near the photophysiological saturation intensity gives improved agreement (Arrigo and Sullivan, 1992), and it is justified on the basis that energy may be required to sustain the extra carbon fixation.

We conclude with a review of organic molecular behaviors in Arctic sea ice, with attention to forms and concentrations which may be capable of altering channel structure (Janech et al. 2006; Krembs and Deming, 2008; Underwood et al. 2013; Aslam et al. 2012a). Pitting, films, gels and a general inhibition of formation are all considered relative to the modeled macromolecular concentrations. The potential is sketched for a full microphysical approach to adsorption and additional phase transitions within the pore network. Our results point strongly toward biological controls on crystal formation, both in bottom ice and during intermediate or upper level blooms in many remote locations. We end with recommendations for a renewed measurement effort, coordinated with global scale simulations of the processes involved. Development of a bi-hemispheric approach is suggested, so that the concepts can also be tested in the Antarctic.

**2. Simplified Physical Model**

Input fields for major physical variables were taken from history files generated by CICE (Hunke et al. 2015) configured in its most recent release (5). The parent code is known for its extended viscous-plastic treatment of rheology, with an elastic wave mechanism introduced allowing for explicit numerics to represent responses to stress. A thermodynamic subunit computes local growth rates of snow and ice due to vertical conductive, radiative and turbulent fluxes. Features critical to brine channel biogeochemistry include a mixing length approach to vertical tracer transport, multiphase porous flow (mushy layer theory), incremental remapping advection in the horizontal and a delta-Eddington multiple scattering scheme. Early applications are discussed relative to general tracer transport in Jeffery et al. (2011) or Jeffery and Hunke (2014). For an example of bottom layer biogeochemistry plus Arctic aerosol precursor simulation, the reader is referred to Elliott et al. (2012). CICE forms the core marine cryological simulator for several climate system models (e.g. Collins et al. 2006; Rae et al. 2015) and also a new Department of Energy code referred to as ACME (the Accelerated Climate Model for Energy).

In preparation for the present offline tests, a standalone CICE run was conducted from 1980 to 2009. Ocean forcings were taken from POP set in the Gx1 configuration within CESM (acronyms refer respectively to the Parallel Ocean Program and Community Earth System Model –e.g. Jeffery and Hunke, 2014; Hunke et al. 2015). The atmosphere overlying is simulated through the CORE II climatology as described in Large and Yeager (2009). The ability of an independent ice model to capture variation in coverage and thickness over multiple decades has been demonstrated by Hunke (2010). Melt ponds were recently been incorporated into CICE and then improved with respect to their numerics (Hunke et al. 2015). In the real polar system surface ponding supports considerable biological activity, but these relatively fresh features will be considered separately and at a later date. We restrict ourselves here to chemistry within the brine.

From results of the standalone simulation, lower tropospheric, ice and seawater temperatures were averaged at multiple levels along with simulated thicknesses of snow and ice cover, over a period of several years near the turn of the last millennium (predating major climate change-driven coverage losses). The data were then further averaged areally, over the biogeographic zones originally defined by Carmack and Wassmann (2006) then expanded by Deal et al. (2011). This simplifies the full (Arctic) polar ice algal challenge, reducing it to ten one dimensional problems. The zones are mapped in Figure 1, and they encompass the latitudinal range of all sea ice types. Results will in general be presented and discussed beginning closest to the subtropics, so that the Sea of Okhotsk is often treated first. By the same token, the central Arctic Ocean is often the last area to come under discussion. For zones ringing the basin at the northern edge of the continents, we begin our analysis at the Greenwich meridian (GIN Seas) and then move eastward.

All biogeochemical simulations begin on the date January 1 in a generic year near the turn of the century. Initial conditions are determined by inserting deep winter nutrient and organic levels into the available vertical pore spaces during that time. For organics we make the assumption that only the humics are present in the period. Fresh macromolecular concentrations begin at negligible levels. Desalination later in the winter or in spring plus an evolution of biopolymer concentrations as blooms occur below the ice are both parameterized within the model mixed layer, in order to reflect measurements. The organic patterns are discussed in Elliott et al. (2014) with regard to the Ross Sea, but results are phase shifted by six months for the Northern Hemisphere. Subsurface observations are in accord (Dittmar and Kattner, 2003a and b). Flooding, which is better documented in the Southern Hemisphere (Ackley et al. 2008; Meiners et al. 2012), may also occur in the North and its onset is computed according to Archimedes Law (Fritsen et al. 1998). Sufficient buildup of snow on the pack may cause sinking with infiltration, particularly when/where the ice is thin. Nutrient and organic concentrations spread onto the depressed ice surface as dictated by data for the local mixed layer.

**3. Biogeochemistry**

The physical driver quantities are distributed in weekly increments over a typical year and also the regional categories of Figure 1, forming multiple fields/matrices of size 10 by 52 (zones by week). This array of values was taken as input into a numerical model for vertical ecodynamics, chemical kinetics and transport within sea ice at all locations. Biogeochemical continuity equations were constructed to represent the time evolution of major ecological and organic chemical quantities. The equations and baseline parameter settings are laid out in detail in appendix materials (A and B), and the model system is mapped schematically in Figure 2. An informal list of tracer groups presented in the appendix notation would be **auto** (all relevant autotrophic organisms encompassing both phytoplankton and ice algae, but often this is merely a renaming), **nut** (the major inorganic molecular nutrient forms including redox states plus iron), and finally **mac** (either fresh or aged-processed biomacromolecules and polymers). Within each grouping elemental ratios can be maintained as desired, but carbon always functions as the major currency. Subtleties include further sets accounting for the nutrient elements in and of themselves, core elements within the various dissolved forms, and macromolecular carbon content. These are required to deal respectively with alternative sources, the computation of limits and successional accumulation. Our scheme is built up from published ice biogeochemical simulators focused on pigments distributions, carbon and sulfur cycling (Arrigo et al. 1993; Lavoie et al. 2005; Elliott et al. 2012). Some contemporary ice algal models already track the detrital dissolved organics collectively (Tedesco et al. 2012), but in our case a premium is placed upon structural resolution. It has become clear from both physical chemical and geographic standpoints that carrying the macromolecular chemistry may hold advantages (Krembs et a. 2011; Underwood et al. 2013).

Arctic nutrients and organics are fed into the bottom of the lowest stratum of the code then permitted to mix through several numerical layers moving upward through ice column. Transport is eddy diffusive but accounts for the effects of porosity in the simple manner described by Vancoppenolle et al. (2010). A general molecular diffusion coefficient of $10^{-9}$ cm$^2$/s is adopted from Lavoie et al. (2005). Laminar layer thicknesses are adjusted to give observed ice internal residence times (Reeburgh, 1984; Lavoie et al. 2005; Vancoppenolle et al. 2010). Mixing length and mushy layer influences are only known to the reduced model indirectly, through the importation of CICE output. Snow cover was used to compute buoyancy of the overall vertical system per Archimedes principle as in the main model, with a threshold for submersion and generation of an infiltration layer that is entirely adjustable. Porosity was calculated based on the evolving salt content, given the standard freezing point depression equilibrium (Arrigo et al. 1992; Vancoppenolle et al. 2010).

Ice algal blooms are triggered in the model through the relaxation of nutrient and light limitation, subject to salinity restrictions outlined in laboratory photo-physiological studies (Arrigo et al. 1992 and 1993). The salinity growth barrier is computed for consistency with the local porosity. Zooplankton are treated in the mechanism as a nonmodelled background entity which skims a constant low fraction of primary production (Arrigo et al. 1993; Jin et al. 2006). This is a typical assumption even in contemporary ice ecodynamics simulations and it will be considered in more detail in the discussion section, since organics are affected directly. Grazing and mortality are routed into a complex network of spillage, assimilation and remineralization pathways (Elliott et al. 2012 and the appendix here). Dissolved carbon of biological origin is present as a refractory humic background (Calace et al. 2001) but with a time dependent superposition of fresh macromolecules due to cell disruption (Elliott et al. 2014). Since the fresh releases were not always sufficient to explain observations, organic exudation was ultimately introduced in several modes in order to provide an additional set of sources. Constant injections and analogs varying with relative light level were both tested extensively.

The most critical tracers in the system include the nutrients nitrate, silicate and also iron –the latter since an eventual goal is transition to the Antarctic (Tagliabue and Arrigo, 2006). The autotrophs whose behavior is simulated are the ice diatoms (usually pennate), generic microflagellates and a collective aggregate of *Phaeocystis*. This ensemble of organisms follows the lead of Walsh et al. (2001 and 2004) in establishing an algal vector applicable in both hemispheres. Labile organics are simply the major classes of macromolecule comprising all biological soft tissue (Benner, 2002; Underwood et al. 2010); the proteins, polysaccharides and lipids. Mixed polymers naturally exist and may well be critical to ice structural effects (Krembs et al. 2011; Aslam et al. 2012a), but they are treated mainly as conceptual combinations of the pure analogs. All biogeochemical quantities are represented in the figure schematic, with details defined by the appendix parameter and equation lists. Our ice algal mechanism derives from the pioneering work of Arrigo et al. (1993) but with Arctic adjustments (Lavoie et al. 2005; Jin et al. 2006), Pan-Arctic extensions (Deal et al. 2011; Elliott et al. 2012) and finally the addition of iron so that the option is available for Southern Hemispheric work (Thomas, 2003; Lannuzel et al. 2010; Wang and Moore, 2012).

Surfactant properties of the biomacromolecules are now given serious consideration along marine bubble, microlayer, spray and primary aerosol interfaces (Elliott et al. 2014; Burrows et al. 2014; Carpenter and Nightingale, 2015). Organics also act as ligands for trace metal retention in sea ice relative to open water (Van der Merwe et al. 2009; Lannuzel et al. 2010). Fresh releases are divided into the generic forms in a classic marine algal content ratio 60, 20 and 20% (protein, carbohydrate, lipid -Elliott et al. 2014). Ice organic studies indicate that in some cases biopolymers influence pitting of the brine channel walls (Raymond et al. 1994; Janech et al. 2006). In other instances, cross or hybrid polymers such as glycoproteins may gel and force changes in tortuosity (Krembs et al. 2011), but for present purposes their existence must be inferred from pure model compounds. Time evolution of the collective fresh organic mass is simulated via release due to the various forms of cell disruption, and then direct exudation as a sensitivity test (Tedesco et al. 2012). Our overall philosophy is to mimic the role of dissolved and colloidal organics within the carbon cycle as a feature of ecodynamic succession (Underwood et al, 2010; Tedesco et al. 2012; Aslam et al. 2012a). Such behavior is often reported in observational time series (Amon et al. 2001; Thomas et al. 2001) and has recently been demonstrated for global sea ice at a statistical level (Underwood et al. 2013). Although the refractory humic acids are treated as a constant background in the water column, like all other quantities they are mixed and supplied dynamically to sea ice from below. Porosity is expected to be a factor determining bulk concentrations.

In more localized ice biogeochemical simulations, the vertical system is sometimes decomposed into a continuous series of thin layers (Arrigo et al. 1993; Vancoppenolle et al. 2010). For present reduced-model purposes, however, we divide into four generic boxes. These are the bottom layer well known in the Arctic (Deal et al. 2011; Elliott et al. 2012), the ice interior which may also be ecologically very rich (Gradinger, 1999; Gradinger et al. 2005; Underwood et al. 2010 -though low temperatures restrict habitat as desalination takes place), plus the freeboard and infiltration layers as defined heuristically based on observations (Ackley et al. 1979; Ackley and Sullivan, 1994; Haas et al. 2001). In some cases the latter two can be difficult to distinguish in the field, and they are relatively rare in the northern hemispheric environment (Melnikov, 1997). Upper level blooms are most often observed and described for the Antarctic but our intent here is two-fold: to explore their potential to exist in understudied portions of the Arctic, and also to prepare for modeling of the thinner, circumpolar ice systems of the Southern Hemisphere. Our four ice sections are given baseline thicknesses of 3, 30, 3 and 3 centimeters (bottom, interior, freeboard and infiltration –(Gradinger et al. 1999; Haas et al. 2001; Lavoie et al. 2005), but these fixed choices are varied in sensitivity tests and permitted to evolve in some individual runs.

Attenuation of downwelling radiation through the layers was accomplished including contributions from snow, ice and total biopigments but in the sense of single scattering using Beer's Law. This follows the precedent set by Lavoie et al. (2005) and carried forward in the aerosol precursor experiments of Elliott et al. (2012). Ice and snow interactions with incoming solar follow parameterizations from physical simulations (Flato and Brown, 1996) and the pigment absorptivities are given in tabular form in the appendix. The tracer transport-continuity equations were discretized and cast primarily into implicit solution forms. Biological growth is extremely stiff numerically so that careful consideration had to be given to operator splitting issues. A strategy eventually settled upon for nutrient uptake is described in the appendices (Vancoppenolle et al. 2010; Elliott et al. 2012). The organics by contrast are long lived in the brine network (Amon et al. 2001; Riedel et al. 2008) so that conservation was achievable automatically, independent of the integration style. Simulations were conducted on a set of laptop computers in the R statistical programming and analysis package, which is publicly available and strongly supported by a worldwide user community. The experiments are designed primarily for the rapid turnover and testing of new features, and ultimately for handoff to full-scale systems models such as CICE. Our Arctic ice algal R package is available from the authors on request.

**4. Data Selection**

Several groups have organized chlorophyll measurements for different vertical strata in Arctic sea ice (Melnikov, 1997; Arrigo, 2003; Elliott et al. 2012). Observations are much more numerous for the bottom layer than for interior or upper level habitats. This is partly but not entirely a reflection of lower level dominance. We distill many of the available data in Table 1, categorizing them by month of the year and the Figure 1 map, then internally. Upper ice habitats in the Arctic are probably restricted in time and space due to reduced snow thickness. Our policy is to lump any measurements which may be relevant to potential freeboard and infiltration layers, though under appropriate circumstances the two habitats can in fact be distinguished (Ackley et al. 1979). Among pack ice organic chemical measurements it is total dissolved organic carbon that is most often reported (DOC), and typical data are included in the table. At perhaps a third of experimental sites the two quantities chlorophyll and dissolved carbon are reported together. The majority of measurements are conducted in the springtime, coinciding with the bloom period and the balance of mild weather with continuing pack coverage. Autumn does not go completely unrepresented, however.

Roughly speaking bottom layer biology appears to maximize at order several hundred standard units of chlorophyll or micromolar of dissolved carbon. A drop in intensity may be discernible moving poleward but the situation is sparse. DOC in the table may or may not correlate with pigments. Determinations specific to the individual organic compound types are also at hand but will be collected later, following the presentation of general results. While not entirely complete, the selected data are fairly representative. This can be seen by comparison with earlier tabulations providing somewhat less detail (Arrigo 2003 or Elliott et al. 2012). Upper level ecosystems are mentioned for the Arctic almost solely anecdotally, for example in the monograph by Melnikov (1997). A working hypothesis is that freeboard and infiltration layers better studied in the Southern Hemisphere (Fritsen et al. 2001; Meiners et al. 2012) occur only peripherally and occasionally in the Arctic.

## 5. Accuracy, Optimization and Sensitivities

Most simulations were conducted at a time step of order one day -short relative to decay constants of the organics but long by comparison with intense ice algal doubling rates in the bottom layer. In the upper levels, growth processes were often delayed by implicit integration. However, they still resulted in reasonable maxima since mass limitations controlled the eventual outcome. Nutrient levels below the pack interface were fixed at observed mixed layer concentrations and were not permitted to relax even if rapid bloom uptake was underway. In other words, instantaneous mixing was assumed in the water column below the ice. This is in common with most of our simulations predating the current model (Jin et al. 2006; Deal et al. 2011; Elliott et al. 2012). In the bottom layer, concentrations tended to be capped by self-shading. But due to the long step size and rich resources upcoming from below, overshoot remained a possibility.

Despite these numerical issues, all major features of the ice algal system were represented realistically. In Table 2, bloom maxima are compared with measurements for the ten regional ecozones. Although the computed primary production is phase shifted in some entries, peak heights are in reasonable correspondence. This is true whether the results are viewed from the ecogeographic standpoint, across latitude or even for the entire polar regime. In the traditional volumetric chlorophyll units of mg/m$^3$, a typical seasonal buildup is in the hundreds. But the physical system can prevent biological activity entirely, for example through the accumulation of thick snow cover. Nonblooming model cells or zones mean low biological activity in the majority of cases documented here. Overall quality demonstrated by the table is sufficient for present purposes, which are merely to investigate ice internal organic chemistry depending on background nutrient cycling. Similar comparisons were conducted with interior and upper layer pigment data, with similar though less robust agreement because observations remain sparse.

Optimization tests were conducted by operating on multiple variables including the dimension of the bottom layer (Elliott et al. 2012), transfer velocities determined from diffusivities (Vancoppenolle et al. 2010) and laminar layer thicknesses (Lavoie et al. 2005). Additionally, dependencies were studied for the main external forcing functions snow depth and surface flooding. The statistical method most often adopted to assess our results was as follows: Bulk chlorophyll figures were log transformed then converted to Mean Absolute Error relative to the data (MAE -Zar, 1984; Press et al. 1996). This calculation allows a quick assessment against observations when values are distributed in a non-Gaussian manner over multiple powers of ten in concentration. Focussing upon maxima from Table 2 while ignoring the time coordinate, convergence was usually well within an order of magnitude at any given location. In other words, peak sizes were adequately represented. This confirms our confidence in proceeding to the simulation of organic carbon chemistry. During the above manipulations, zero or decaying concentrations are limited to a low of 1 mg/m$^3$ to approximate or synthesize the analytical limit of detection.

When a log MAE analysis was tightened to include time, results became more difficult to interpret. Figure 3a reveals that offsets in the maxima are likely attributable to a switch-like influence of the regional snow depth. Ice algal growth was under light limitation for much of the integration period. In ecogeographic zones with the greatest precipitation buildup over winter and into the spring, phase mismatches are introduced which bias the model in a nearly log-uniform manner. A lower limit for bulk ice pigment detection was again taken as the logical baseline for construction of the statistics. Cumulative MAE builds up where snow coverage is thickest. But since observations are sparse even for the bottom layer, there are also zone sequences of little or no change. Divergence between model runs with varying bottom layer thickness was minimal, and attributable mainly to the expected dilution. Quality was difficult to distinguish under these circumstances. We conclude that true optimization of the bottom layer ecodynamics should/must be deferred until our mechanism can be implemented in the full ice system model. Closer examination of the interaction of snow depth with biological activity is definitely a priority (Lavoie et al. 2005; Jin et al. 2006).

So far our discussion of sensitivity has been exclusive to the bottom layer. Upper habitats are so rarely observed in the Arctic that even their existence can be questioned. Although coloration of the freeboard and infiltration level is commonly reported in the Southern Hemisphere (Ackley and Sullivan, 1994; Ackley et al. 2008; Haas et al. 2001), Melnikov claims that top level biology can actually be superseded by the deposition of wind blown organisms (1997). His arguments, however, are specific to coastal Siberia. The Laptev and East Siberian Seas are in fact relatively high in latitude, and snowfall is moderate. Infiltration may be more common for thin ice of the peripheral Okhotsk, or else along the southern edge of Bering and Canadian systems. In separate simulations focused on biology high in the pack, we varied the (Archimedean) threshold for flooding in order to demonstrate that infiltration layers can be eliminated. Upper habitat chlorophyll-days are integrated, summed then displayed in Figure 3b. Once again presentation is in an accumulating mode, and two points become apparent immediately -sensitivity to the flood parameter is extreme and coordinated measurements are sorely needed. As in the case of the atmospheric interactions, inundation physics will be critical to comprehension of pack biogeochemical systems (Arrigo et al. 1997; Fritsen et al. 1998; Haas et al. 2001).

In further test series, the long initial time step setting was altered both upward and downward. Changes in chlorophyll concentration ranged over almost an order of magnitude in each direction. Under step lengthening, serious kinetic instabilities became apparent despite precautions taken for the maintenance of atom conservation. Pigment levels oscillated in several of the layers we defined, due to tradeoffs of excessive shading and ammonia-driven remineralization followed by rebound blooms. Downward step size experiments yielded a certain amount of variation in the height and timing of Table 2 maxima, but since mass and light constraints were specifically designed to bracket the measurements these biases are not serious. We were prevented by computational restrictions from pursuing the matter further. The reduced model is momentarily both targeted for, and confined to, laptop computers. Again we advocate rigorous optimization, when possible in the full CICE code.

**6. Baseline Results: Inorganics and Biology**

Our presentation of time evolution plots begin with nutrients and continue through the primary producers. An Arctic setting for the experiments means that nitrate is most often limiting, and so this is the only geochemical driver quantity included. Although silicate restrictions on diatom growth are conceivable (Walsh et al. 2004), in our simulations such instances are rare. Dissolved trace metal concentrations were set to high levels reflecting strong Northern shelf and riverine inputs (Maeda, 1986; Aguilar-Islas et al. 2007). Model parameters as listed in Appendix B dictate the primary organisms involved, and these are set to match better known pelagic ecodynamics so that coupling with open water will be facilitated. The choices are typical for the Bering and Pacific Arctic entryway (Walsh et al. 2004). Detailed observations of cell densities within the pack are called upon for initialization (Gradinger 1999; Fritsen et al. 2001). Competitive exclusion dominates in our output, but the real situation is generally not so extreme. Microhabitats are thought to enable coexistence (Hutchinson, 1961), but are not included here.

Starting with the Sea of Okhotsk as a light replete case, we repeat the integration, sensitivity and optimization exercise moving northward. But in this section analysis falls on the Okhotsk and then three further, critical biogeographic zones -the Chukchi and Beaufort since they are relatively rich in dissolved carbon data (e.g. Krembs et al. 2002; Uzuka, 2003; Gradinger et al. 2009) and the central (polar) subdivision since it defines an extreme. Bulk concentrations are dealt with exclusively since in most cases measurements are performed on melted core slices (see Appendix A for the conversion to brine intrinsic levels). A strategy of presenting bulk base 10 logarithms is maintained throughout, since values which must be intercompared and explained range over orders of magnitude. Model output is transformed in this manner then plotted against week number, over the year of simulation. No attempt has been made to continue into or through a second winter season –wrapping will be better suited to a supercomputing environment and the complete model.

Results for certain inorganic or biological quantities are presented initially from the baseline in Figure 4. We now proceed to interpretation, moving from the upper left subpanel and then upward through the ice column meaning right and downward over the plot set. Rapid equilibration into the bottom is expected for mixed layer nitrate. The diatoms bloom rapidly in the Okhotsk since sunlight is already at hand, but the rate is nutrient flux limited (Elliott et al. 2012). Siliceous organisms are ecologically dominant since they go ungrazed and retain their position in the skeletal matrix (Lavoie et al. 2005). (Relative) immobilization strategies probably involve excretion of the organics (Krembs and Deming, 2008; Underwood et al. 2010). An initial flagellate population decays due to outbound mixing. Once the diatom bloom tops out, nitrate levels are maintained near those acting as a source from the mixed layer, but with porosity at about one half (Arrigo et al. 1993). All Sea of Okhotsk concentrations crash at the point where CICE indicates melting above, since rapid percolation and flushing are assumed (about week 17 -Vancoppenolle et al. 2010; Appendix A). A rebound of both nutrients and the diatoms is supported in the summer since the reduced model allows ice to linger briefly after cessation of the melt (weeks 30 to 31). Persistence occurs because the model turns flushing on and off through an adjustable threshold meant to mimic a shift from floe runoff to vertical percolation (Eicken et al. 2002). It remains to be seen whether this brief seasonal feature can be reproduced in the full code or even whether it represents reality in the environment. We are not aware of biogeochemical observations during the relevant period -and they would no doubt be difficult to obtain on thin ice. Pack reformation in the fall begins the bottom layer ecology anew -note that autumn blooms have not gone unreported in the literature.

The interior ice column communicates with its active neighboring layers in the Okhotsk, but internal porosities are low and restrictive of biological activity (Arrigo and Sullivan, 1992). The potential for a bloom is in fact suppressed until spring. The diatoms are not permitted to cross from the bottom matrix to internal channels in our simulations, and so do not make a contribution to interior carbon. A summer nitrate return peak reflects nutrient transport attributable to the mixed layer, but concentrations are low and no biology ensues. At the freeboard there is a strong mass limited bloom which is triggered by increased porosity, but this feature is terminated instantly by the melt after only a few weeks. Net algal levels are orders of magnitude less than those of the bottom layer and smaller photosynthesizers now dominate (Gradinger, 1999; Underwood et al. 2010). For simplicity Phaecystis populations are not shown alongside those of the flagellates, but since many of the growth parameters are in common this class is also competitive (Tison et al. 2010). The post-melt recurrence feature is absent from our model freeboard at low latitudes. Nutrients have been flushed from the upper ice but cannot be resupplied from the ocean below in a timely fashion, due to distance and porosity influences.

In the lower right of Figure 4, results suggest an infiltration habitat coinciding closely with the spring-centered bloom just described. This activity is positioned immediately above the freeboard in the model, and would likely be difficult to segregate (Ackley and Sullivan, 1994). Again concentrations are mass limited, this time by the incoming nitrate content of the local seeped/flooded brine injection, so that the levels of activity are very low relative to the ice bottom. Upper level biology may be more intense in the Antarctic for a deceptively simple reason –mixed layer nitrogen levels are significantly higher there. The Southern Ocean is famously categorized as the largest of all global HNLC zones (High in Nutrients but Low in Chlorophyll; Fung et al. 2000). Southern Hemispheric sea ice seems to be capable of scavenging iron during frazil stages, so that pelagic trace metal limitations may not apply within the pack (Lannuzel et al. 2010; Wang et al. 2014). As discussed for Figure 3b, existence of Arctic infiltration layers is speculative, and they must be considered a tentative outcome of the present set up.

Figures 5 through 7 display analogous results for more northerly regions. Many of the above comments could be repeated relative to the Pacific Arctic sector, but subtle variations are apparent. In the Chukchi Sea, snow reflectivity prevents photons from penetration and this delays the diatom bloom. But infiltration sets in early, and light is able to penetrate at least this far through the system. A mass (mole) constrained level is therefore established and preserved at the top of the ice column. In the Beaufort the bottom bloom is even more intense than at lower latitudes, but it is shifted significantly and again snow cover is responsible. Suppression of the supporting nitrate is seen during an especially intense growth period as a "bite" taken out of the time evolution profile (weeks 9 to 13). The infiltration layer however is completely absent in this sector, again because the snow data provided by CICE are insufficient to cause flooding. At the North Pole biological activity shuts down but not entirely (M. Levasseur, unpublished data from Sharma et al. 1999 cruise). A modest infiltration layer bloom builds just prior to the melt. This may be a real but unnoticed feature, since the remote central Arctic Ocean is so little observed.

**7. Baseline Results: The Organics**

We display time evolution of concentration for three sample organic tracers. Background humic acid mixes or freezes in at a constant level reflective of the seawater source, and it is then modulated in bulk by the local porosity. Since this carbon pool is considered refractory (Malcolm, 1990; Calace et al. 2001; Benner, 2002), inputs and channel tightening are the only factors involved. By contrast the proteins and polysaccharides are injected even at a minimum by grazing-mortality processes, as determined by the network of routings (Appendices A and B). The compounds are long lived relative to the bloom scale so that there is a tendency to accumulate during local biological activity (Aslam et al. 2012b; Underwood et al. 2013). Downward transport may exhaust them quickly from bottom ice (Lavoie et al. 2005; Elliott et al. 2012) but at other levels month-long decay is possible and sometimes apparent. A rich oceanic dissolved organic chemistry may in fact be driven independently by under-ice blooms (Alexander, 1974; Arrigo et al. 2012). The effects are parameterized here through variable mixed layer concentrations in springtime, set highest for the longer lived polysaccharides (Elliott et al. 2014).

Results for the warm ice system of the Okhotsk are shown in Figure 8. In the bottom ice layer, concentrations of pure macromolecules are maintained near micromolar primarily by diatom mortality. Grazing is not permitted in this case and initial flagellates have long since decayed (Figure 4). After ramp-up corresponding to the early bloom, concentrations attain a steady state balancing the mortality inputs with downward mixing into the sea. Proteins are in greater abundance since we assume direct release via cell disruption and a typical composition for autotrophic organisms is fixed at the global average (Parsons et al. 1984; Wakeham et al. 1997; Appendix B). Notice that following the melt and prior to disappearance of the pack, the organic source signature is reversed –the thin and intermediate thickness curves have crossed. At this point carbohydrates are in excess, corresponding with relative oxidation rates below ice. The concentration switch holds for both bottom and interior, because the two are in fairly close communication with the solutes of the mixed layer. At the freeboard level, channel tightening prevents growth and flagellate mortality raises organic concentrations only slightly until biological systems open up during the spring (roughly week 13). From this stage, grazing emissions bolster organic geocycling. The infiltration layer enters its bloom phase immediately since it is close to the overhead light source. The biomolecules track their top level algal producers closely.

In all Okhotsk cases, fresh organics are generated mainly by mortality until photosynthesis begins. Fractionated grazing and spillage processes then come into play, and the dissolved carbon is always subject to heterotrophic (bacterial) removal on a one moneth scale (Appendix B). Concentrations supported by baseline ice algal behavior in this region are of order micromolar or less at most times of the year. The summer rebound approaches a value of 10, but is attributable to organisms operating independently in the water column. Freeboard and infiltration peaks are brief and corroborative observations are lacking. Results for all other biogeographic zones are consistent with these interpretations. In the Chukchi (Figure 9), snow prevents an early bottom layer rise but the bloom scenario is replayed in autumn. Mixed layer polysaccharides-proteins enter as a virtual step function (week 20) since the melt takes place only after they make their parameterized appearance from below. Clearly this situation can be improved/smoothed in coupled ocean-ice calculations. Infiltration layer buildup is substantial, approaching ten micromolar of total fresh biopolymeric carbon. The Beaufort and central scenarios follow in kind, with the former resembling the Okhotsk and the latter exhibiting little activity.

The protein and polysaccharide bins track one another closely in our simulations because they are released and removed in constant ratios. During real world succession more complex relationships are observed (Thomas et al. 2001; Thomas and Papadimitriou, 2003). Furthermore the full complement of biomacromolecules necessarily includes mixed functionality polymers that are combinations of idealized compound classes represented here (Benner, 2002; Underwood et al. 2010; Krembs et al. 2011; Aslam et al. 2012a). Lipids are highly surface active at the water-air interface since they are insoluble (Elliott et al. 2014), and this class is roughly as abundant as the carbohydrates in our calculations. Aliphatic bubble coatings may thus be anticipated in the monolayer sense. Model-generated lipid profiles merely resemble those of our proteins. The latter, however, are much more often discussed in the literature for their potential ice-structure altering capabilities (Raymond et al. 1994; Janech et al. 2006; Krembs et al. 2011; Underwood et al. 2013). We thus elected not to portray lipidic carbon content in our plots, even though the chemistry is just as detailed (Appendices A and B). Generally the sum of network carbon excluding refractory humics tends to be of order micromolar or much less in our baseline results. Observations and the levels required to match experimental demonstrations of structure change are considerably higher (Table 1; Krembs et al. 2002 and 2011). Hence we turn now to the process of exudation.

## 8. Sensitivity Tests on the Organics

Observed Arctic sea ice carbon concentrations often rise significantly above the levels simulated thus far. Proteins, polysaccharides, humics and total organic carbon have all been detected at significantly greater than unit micromolar in the bulk (e.g. Calace et al. 2001; Amon et al. 2001; Thomas et al. 2001; Krembs et al 2011; Underwood et al. 2013). To this point in our development, it has been unnecessary to distinguish the physical state of brine organics. We have so far cited only measurements of the DOC (Table 1), components of which must reside in solution as a matter of definition. Filtration is applied in the laboratory and particles exceeding some small number of microns are readily separated. But as individual polymer types enter the analysis, viable data are increasingly fragmented. Simultaneously both the pore chemistry and laboratory analysis become more involved (Underwood et al. 2010; Aslam et al. 2012a). From this point we will follow the conceptual macromolecular dynamics model and definitions of Underwood et al. (2010 and 2013) to organize comparisons. Whether release into the channel network occurs via algal membrane disruption or some alternate means, biomacromolecules are assumed to pass through an initial dissolved state representing a spectrum of molecular weights - but they are sometimes observed only much later in aggregate form. The possibility of discerning phase transitions will in fact constitute a direction for future applications. But we propose from this stage to combine filtered material with filtrate in order to obtain the maximum amount tracer information at a molecular level.

Even after such conceptual lumping the situation remains complex, as is typical for the environmental organics. Injected polymers may subsequently be degraded to oligomers or even monomers (Thomas et al. 2001; Benner, 2002; Underwood et al. 2010). The internal reconstitution of humic substances is yet another possibility (Calace et al. 2001). But our fundamental goal is to understand mass redistribution during succession, so that a collective approach should be sufficient for the moment. Organic matter is thus considered regardless of its analytical phase state, and data availability increases. Detailed sorting into colloidal or condensed size distributions will be undertaken only later. Not incidentally in our data management process, living matter has been classified in the usual way as particulate organic carbon or POC. This is of course essential to a recognition of freshly released biochemicals since they are highly concentrated inside primary algal sources. The appendix mechanism can now be thought in the following manner: Macromolecules emanating from producer cells may well be dissolved initially, but since condensation is not yet considered we remain agnostic toward the evolving physical state. It is our intention to introduce polymer kinetics and gel formation processes in next generation models, relative to both ice and open water biogeochemistry.

Several chemically resolved studies of the Northern Hemispheric ice content are summarized under these restrictions in Table 3, and a selection of analogous Antarctic data is appended in the last row since the values are in accord. Total DOC is typically present in the melt at hundreds of micromolar regardless of bonding structure (Underwood et al. 2013). This result does not conflict with a high refractory background shown as the porosity-modulated humic acid profile (Calace et al. 2001). But a strong supplementation is suggested. Other measurements are made by means of chemically-specific techniques such as staining, or liquid chromatography in conjunction with absorption spectroscopy. Comparable levels are often identifiable as protein or carbohydrate (Amon et al. 2001). So far our model underrepresents these features of the polar biogeochemical system by orders of magnitude. We therefore experiment with forced organic release in addition to cell disruption already computed along several pathways. Also provided in the appendices are exudation rate constants which are now enabled for sensitivity testing. Like the more general cellular detritus, exuded species are presumed to enter in their global average ratios to biomass (Wakeham et al. 1997). This is clearly a gross simplification, and the saccharides do tend to attract more attention as a major component of EPS (the extracellular polymeric substances –Krembs et al. 2011; Underwood et al. 2013). But the assumption that proteins should be included in the mix is not inconsistent with an excess of total fresh organics. Extra carbon is often observed over and above the sum of saccharides and humics (Reidel et al. 2008; Underwood et al. 2010). Note that the exudation process differs from cell disruption in that it can be decoupled from ice algal grazing or death.

During initial simulations, the fresh biopolymers were injected from all extant primary producers at a fixed rate of one tenth of their carbon content per day, regardless of local light or nutrient limitation. This pace is likely sustainable for order weeks, since it represents a small proportion of the maximum marine photosynthetic rate (Eppley, 1972; Arrigo and Sullivan, 1992; Lavoie et al. 2005). Next the emissions were limited to periods when the light intensity exceeded some set proportion of photophysiological saturation ($I_s$ -Arrigo et al. 1992 and 1993), since it must be admitted that carbon fixation could require an external energy source. The threshold was graded downward until significant further increases could be documented, and the shift took place at a few tenths of the saturation intensity. In this case the release rate was set to one per day, roughly equivalent to an ice algal growth maximum.

Given a constant background emission at the slow rate of one tenth per day, we found that intermediate intensity ecosystems were only able to sustain up to 10 micromolar of combined protein and polysaccharide, hovering below the background of refractory polymers. This was true for the Sea of Okhotsk and Beaufort at all levels where sea ice existed, and in the Chukchi for the infiltration layer. In the central Arctic, levels of the fresh organics were raised one to two orders of magnitude but were still lower than measurement data and fell well below unit micromolar of carbon. For all ecogeographic zones, profiles resembled those of the organic time evolution figures already discussed. This is because mortality and grazing releases differ in concept but not rate under the mechanism. Light-linked results, however, contrasted dramatically both in terms of trajectory and average concentrations. Organic intrusions were strong and dynamic where they were switched on. In all four of our standard scenarios as shown in Figures 10 through 13, dissolved concentrations now fall in line with Table 3 data for many layers and locations.

In Figure 10 the resemblance of initial Okhotsk output to Arctic studies such as Krembs et al. (2002) or global patterns as collated by Underwood et al. (2013) would be striking if the results were general. Taken together the biomacromolecules carried in this simulation total to hundreds of micromolar of ice-internal carbon (sum protein, polysaccharide and humics noting that lipids go unplotted). In fact the Okhotsk result is reproduced elsewhere with a phase shift following solar angle, except where excess snow cover prevents primary production. Order ten to one hundred micromolar of fresh biopolymeric carbon becomes the rule during in season, and it may be thought of as additive on a humic background. Upper level systems often exceed hundreds of micromolar of the macromolecules. In the Chukchi Sea (Figure 11), snow data ingested from CICE still forestall both light penetration and photosynthesis in the lower ice. Equilibration with the water column remains dominant here, as indicated again by a crossing of protein and polysaccharide contours. But transition to one tenth light saturation is readily discernible at the freeboard, and both this level and the infiltration layer can now support tens to hundreds of organic micromolar. Upper levels bloom in other locations as well. In Figure 12 for example, the Beaufort behaves as a combination of the previous two lacking only infiltration biology. Even in the polar sector where biological activity is expected to be minimal (Sharma et al. 1999), strong pulses of organic activity are apparent just as the saturation radiation threshold is crossed (Figure 13). The series of model adaptations simulating continuous then light-directed exudation can thus achieve substantial agreement with the data (Table 3).

## 9. Summary and Discussion

Macromolecular chemistry supported in brine channels, by ice algal growth operating under extreme temperature-salinity conditions, probably influences physical pack structure up to regional scales. Organic polymers exhibit pitting, gelling and salt retention properties which tend to organize nutrient distributions in the vertical (Raymond et al. 1994; Janech et al. 2006;

Krembs et al. 2011; Aslam et al. 2012a and b; Underwood et al. 2013). Such features of the physical chemistry may allow organisms to maintain position in the ice column relative to seawater resources, or within the rarified polar light field (Ackley and Sullivan, 1994; Lavoie et al. 2005). Simultaneously however, biology alters thermodynamic and mechanical properties of the most reflective system currently countering the greenhouse (Ackley et al. 2008; Lengainge et al. 2009; Underwood et al. 2010; Krembs et al. 2011). We have simulated production and distribution of sea ice biomacromolecules throughout the Arctic environment, in a reduced model with four habitat layers stacked and sorted according to biogeography (Arrigo et al. 1997; Fritsen et al. 1998; Carmack and Wassmann, 2006). The mechanism is an extension of several earlier works focused on the bottom layer (Jin et al. 2006; Deal et al 2011; Elliott et al. 2012), and it can be readily implemented for global brine networks in the CICE code (Hunke et al. 2015). Biogeographic zones for which ice algal growth and organic injection are tested include all ecosystems of the Arctic as shown in Figure 1. Detailed equation and parameter lists provided in the appendices.

Validation focuses of necessity upon the more numerous data available for rich bottom layers (Table 1; Arrigo, 2003; Elliott et al. 2012). For multiple sun-synchronized chlorophyll peaks at the water-ice interface (Table 2), blooms are represented to well within an order of magnitude. Areas of maximum and minimum activity are often accurately portrayed (contrast high biomass with "no bloom" zones). Cumulative time-dependent error statistics were compiled and evaluated, but the results in this case are less concrete (Figure 3a). Snow variability confounds more sophisticated merit functions, such as mean absolute error of the log transformed output. Prediction of intermittent upper level habitats is highly sensitive to the ratio of snow and ice thicknesses, since this is the main determinant of floe depression (Fritsen et al. 1998; Haas et al. 2001; Ackley et al. 2008). Upper level biology is only rarely observed in the Arctic (Melnikov, 1997; Gradinger, 1999; Gradinger et al., 2005). The existence of intense ice algal production near the snow interface remains to be verified.

Standard spring blooms appear at the bottom of the model pack in baseline outcomes, radiation attenuation by snow and ice permitting. Close communication from seawater to the skeletal and interior layers fuels a continuous influx of nutrient, which often restricts growth in Arctic ice. Lower levels of biological activity are mainly light limited and all systems are terminated by the melt (Lavoie et al. 2005; Jin et al. 2006; Elliott et al. 2012; Figures 4-7). Interior ice is usually impoverished despite mixing from below, since small porosities and high salinities restrict photosynthesis until the seasonal warming (Arrigo and Sullivan, 1992; Arrigo et al. 1993). Upper levels carried by our model include both a freeboard and an infiltration layer. These habitats are better known from Southern Ocean studies (Ackley and Sullivan, 1994; Haas et al. 2001; Ackley et al. 2008; Meiners et al. 2012) but nonetheless we include them since they may be peripheral in the Northern Hemisphere and will be needed for the Antarctic. At the freeboard, biological activity once again depends on expansion of the pore structure with rising temperature, while infiltration is simulated as a function of snow loading (Melnikov, 1997; Arrigo et al. 1997; Fritsen et al. 1998). In the latter two strata, blooms are delayed by porosity and flood threshold constraints then truncated early by the melt.

In baseline runs, fresh macromolecules are released exclusively by cell disruption. They pass from the algal pool into ice brine channels and are set against a background of recalcitrant humics (Calace et al. 2001; Van der Merwe et al. 2009; Elliott et al. 2014). Levels of recently synthesized biopolymeric carbon reach only of order one micromolar (Figures 8 and 9) –likely insufficient to support alterations to pack crystalline structure (Raymond et al. 1994; Riedel et al. 2008; Krembs et al. 2011). Sensitivity tests are therefore conducted with exudation superimposed. This new form of release is represented as targeted injection from intact cells. It occurs over and above the flow from mortality or grazing (Appendix A). Constant emissions at a fraction of maximal growth were only partly effective, but given a light requirement agreement with dissolved organic data is obtained. For periods bracketing bottom layer blooms or else intensification of upper ice biology, hundreds of micromolar bulk organic carbon are generated (Figures 10 and 11). Some notable features of our simulations cannot yet be verified, because most Arctic data are available only for established sites (e.g. Alaska and the Mackenzie basin; Riedel et al. 2011; Krembs et al. 2011). A great deal more field study is required before the organic chemistry of sea ice can be completely understood. However, we cannot currently dismiss the possibility of widespread biological control on pack crystal structure. Based on our results, the most interesting effects are likely associated with spring blooms around the Arctic rim with an occasional extension into summer ice. Further investigations are definitely warranted.

Organics likely imply not only ice restructuring capability but also surface activity, adhesion (which may sometimes be irreversible) and colloid formation. The appendix mechanism deals only with basic phases –ice taken as a solid plus internal aqueous solutions. A series of simplifications is thus necessary, even as we prepare to simulate more complex pack physical chemistry. It is tacitly assumed that algal macromolecules enter the brine initially as solute (Underwood et al. 2010). Later interactions with interfaces and with each other are ignored. This serves as a convenient startup expedient, but it also provides an appropriate springboard for next generation work. Entire families of protein and polysaccharide are known to adsorb to the frozen surface, blocking solid deposition in a noncolligative fashion (Walstra, 2003; Janech et al. 2006; Aslam et al. 2012a; Underwood et al. 2013; McNeill et al. 2012). Total area inside the brine network may be extreme, approaching a square meter per kilogram (Golden et al. 1998; Krembs et al. 2000; Light et al. 2003). A single monolayer of polymeric adsorbate could be sufficient to generate tens of micromolar of apparent dissolved carbon after the analytical core-slice-melt sequence (contrast Thomas et al. 2001 with Elliott et al. 2014). It is possible that not only the diatoms but also their detrital macromolecules are retained within the bottom layer, resisting outward flushing pressure from drainage.

We confirmed the potential importance of adsorption through sensitivity tests in which proteins were transported by exact analogy with siliceous organisms (Lavoie et al. 2005; Elliott et al. 2012; appendices). Macromolecular mixing was artificially limited to a single direction from ocean into the pack. Our cell disruption routings were sufficient to trap carbon concentrations exceeding those of Figures 8 and 9, at least with reference to the upper left panels. But in central or upper sea ice layers with slower export, differences were minimal. Hence exudation will likely be required in any case. For simple adsorptive retention to contribute exclusively in the bottom layer, it must be postulated that algae exude specifically within the interior. A medium term goal will be to evaluate accommodation (at solid walls) versus rapid expulsion from the bottom layer. Detailed polymer kinetics will be required over and above the mechanism so far offered (Davies and Rideal, 1963; Birdi, 1989; Underwood et al. 2010). We believe that ice biogeochemistry models must soon advance beyond bulk tracking to incorporate rigorous comparisons with adsorptive, micellular and colloidal equilibria. Interfacial and phase transitions should at some point be simulated dynamically. Only then can the gels so often observed be captured in detail (Krembs et al. 2002 and 2011; Underwood et al. 2010 and 2013). All these features of the ice organic chemistry are under study for future introduction into CICE. The box diagram of Figure 2 will necessarily become much more involved.

Our approach can also be interpreted from the standpoints of the individual organic species. We begin with humic acid since it forms a stable backdrop. The term "humic" is used here in the usual sense to denote mixed detrital biopolymers resistant to enzymatic degradation (Malcolm, 1990). We postulated as a starting condition that such molecules are formed remotely in seawater then entrapped during frazil formation (Calace et al. 2001). Since the heterogeneous compounds are unreactive, we treat them at a constant background level falling in the center of the observed concentration range (Table 3). Little variation occurs except for dips tracking temperature and porosity. In fact however, Arctic humic carbon will vary with location and should be greatest where there are riverine or shelf/slope contributions (Dittmar and Kattmer, 2003a and b). The lower two boxes in our numerical model merely maintain a constant, transport-driven equilibrium with ocean water since we do not incorporate geographic variations in this particular carbon pool. Refractories are usually not discussed in relation to ice structure (e.g. Raymond et al. 1994; Riedel et al. 2008; Krembs et al. 2011). Their hydrogen bonding capabilities are probably too irregular for the matrix to be affected. But they encompass many of the familiar organic functionalities –heterogeneous polymers are in fact built up from bits and pieces of the pure compounds simulated here, or else from hybrids such as glycoprotein. Recalcitrants may therefore play into ice-iron chelation and net geocycling of carbon in the Southern Ocean (Van der Merwe et al. 2009).

Macromolecular compounds secreted directly by polar marine organisms include all the classes represented here as pure compounds, and their physical chemistry has recently been reviewed in sources such as Krembs and Deming (2008) or Underwood et al. (2010). The ice algal organics can in fact be placed in a more general, global biological context. Across the biosphere, microbes release long chains of carbon to form coatings around individual cells for protection, or else to establish consortia and accumulate in films. The divalent cations calcium and magnesium are prevalent in seawater, so that electrostatic bridges form linking carbohydrate strands into colloids and ultimately gel particles (Chin et al. 1998; Verdugo et al. 2004). These pathways must also be included in at least some future work. Ice algae undergo extreme seasonal thermohaline stress (Aslam et al. 2012a and b). Additionally they experience highly nonstandard pseudo-phase transitions including channel locking and the formation of metastable minerals (Golden et al. 1998; Arrigo et al. 2003; McNeill et al. 2012). It is not surprising that sea ice organisms are notable participants in exopolymer generation. Measurements of the fresh compounds are even more sparse within sea ice than for the biota itself, but so far the exudation hypothesis is well supported (final figures versus Table 3; see also the global compilation Underwood et al. 2013). Total organic carbon in the brine may reach hundreds of micromolar during the growing season. Saccharides are most often seen as secretions, and they are typically considered the dominant type (Thomas et al. 2001; Riedel et al. 2008; Aslam et al. 2012b) with monomers and oligomers sometimes distinguishable (Herborg et al. 2001; Underwood et al. 2010). But proteins are measured at high concentrations as well. Along with their organic polymeric hybrids they have long been associated with crystal alterations (Raymond et al. 1994; Krembs et al. 2011; Underwood et al. 2013).

Our model of organic processing captures several of these molecular-level features of the ice chemical system. The mechanism has been designed to superimpose biopolymeric carbon of recent origin upon preexisting humics. All key functional groups are thus represented, from amino acids to carbohydrates and even extending to lipids. We allow only the emission of pure (idealized) macromolecular forms, but equations could be readily adjusted to account for hybrids. We track only carbon and set aside any accounting of the nitrogen in proteins, but Appendix A outlines methods which might be used to conserve multiple elemental currencies. Per our baseline parameter settings, proteins, polysaccharides and the lipids are released from marine autotrophic cells in a ratio of about 60 to 20 to 20 percent of carbon (Parsons et al. 1984; Wakeham et al. 1997; Benner, 2002). Thus all compounds will be present in brine channels in any given growth-graze situation. Proteins are more abundant because they serve multiple roles inside the cell –as information carriers, catalysts and even in a rigid intracellular engineering capacity. But observations in open waters tell us that marine polysaccharide concentrations usually exceed those of the proteins or lipids. The reason is that carbohydrates have a longer residence time -they are less labile when freely dissolved. This is reflected in the midyear cross-over plots, as dissolved organics mix upward from below into several model locations. Sparse Arctic ice data seem to point to an excess of the sugars (Table 3). This shift may be real and may have a similar explanation, but we elect to maintain the excess protein ratio as an initial, practical expedient.

Several lines of evidence suggest that dominance of saccharides is partly an artifact. Amino acids have been observed streaming from land fast bottom ice at high concentration (Arrigo et al.

1995). Protein-carbohydrate hybrids are well known inside the marine algal cell (Parsons et al. 1984) and are implicated in the most recent studies of ice salinity alteration (Krembs et al. 2011). Total dissolved organics are often reported in excess of extracellular carbohydrates (Underwood et al. 2010). We chose here to maintain proteins at their global average abundance for input fluxes despite the fact that they are less often measured. Where total dissolved carbon exceeds micromolar in our simulations, the proteins and polysaccharides can be conceived as a proxy mix. Moreover they have been given a constant removal rate due to a lack of compound-specific kinetic data. Taken together or with carbon concentrations summed, the model biopolymeric content indicates that organic macromolecules are often present at high concentrations in Arctic ice. The quantities computed closely match those leading to pore geometry reconfigurations, whether in the laboratory or in the field. For example, Krembs et al. (2011) reported significant influence on brine volumes and tortuosity from 100 to 1000 micromolar carbon, and point to heat sensitive glycoproteins as the culprits. In future simulations, we intend to refine the chemical spectrum, improve decay schemes and couple to critical micelle thresholds in order to elucidate such issues.

Ecological succession proceeds on a seasonal scale inside of ice, and the timing of some organic concentration peaks should be delayed as a consequence. Fresh proteins, polysaccharides and hybrid compounds flow independently into the channel network and are degraded there by heterotrophic bacteria in order weeks (Thomas et al. 1995; Amon et al. 2001; Krembs et al. 2002; Riedel et al. 2008). The heterotrophic bacteria themselves then inject extracellular polymers (Aslam et al. 2012b). In our baseline mechanism, only a nominal phase lag could be built into the reaction list relative to primary sources. Production of the biopolymers by cell disruption must for the moment remain fractional and fixed, skimming from the general autotrophy at a constant rate. The organisms themselves are removed with generic time constants set at values similar to those of the macromolecules (appendices), and therefore phasing is difficult to discern in Figures 4 through 9 where it should be most apparent. In part this impression can be attributed to weaknesses in the graphics. The plotting step has been set at one week so that an entire year can be displayed, and brine channel chemical profiles are handled separately from the ecology.

The logic of succession, however, suggests two closely related criticisms -both grazing lags and biomass conversions may be underestimated since the zooplankton are not explicit. In other words, cell lysis may be slaved too tightly to the source profile. Following a long-standing tradition in ice modeling, secondary consumption has been treated solely as a fixed proportion of growth (Arrigo et al. 1993; Lavoie et al. 2005; Jin et al. 2006; Elliott et al. 2012). The usual justification is that understanding simply remains inadequate for higher trophic levels. Plus it is sometimes argued that consumption rates are reduced inside the brine due to lack of access (Lavoie et al. 2005). Data are not entirely lacking, however, to describe the sea ice microfauna (Gradinger, 1999; Gradinger et al. 2005) and open water modeling schemes may be adaptable to the task (Fasham et al. 1993; Sarmiento et al. 1993; Moore et al. 2002 and 2004; Tedesco et al. 2012). Dynamic grazing will be introduced in future experiments, but presently we can neither exclude nor explore the competition with exudation. Disruptive macromolecular release may rise by an order of magnitude if net fractionations are enhanced by realistic zooplankton. Decoupling is by contrast a matter of definition for the light-regulated injections. The few relevant measurement sets are too coarsely resolved to settle the matter (e.g. Riedel et al. 2008).

A major result of the current work is that ice algal observations and modeling converge at the Arctic scale, supporting crucial new roles for brine organics. This diverse suite of molecules may exert control over crystal structures, pore volumes, vertical stratification and nutrient retention within much of the global sea ice domain. Our conclusions clearly apply across multiple sympagic ecosystems surrounding the northern Pacific and Atlantic Oceans and extending into the central basin. As northern pack ice thins and becomes more seasonal in the next few decades, outcomes which are peripheral in our simulations will become prominent features. For the moment superstructure flooding is less often reported in the Arctic, but in the Southern Hemisphere upper habitats are common (Ackley and Sullivan, 1994; Arrigo et al. 1997; Haas et al. 2001; Meiners et al. 2012). The model outcomes suggest that multiple strata of activity will encroach upon the pole. Direct links to the traditional carbon cycle can be imagined as well, and in particular this will be true in the south. Since Antarctic sea ice strips iron from the Southern Ocean water column (Lannuzel et al. 2010), brine channel organics may function there as a factor in trace metal storage (Hassler et al. 2011). Filtering into the pack may in a sense nullify the iron limitation for which Southern waters are well known (contrast Fung et al. 2000 with Thomas, 2003; Wang and Moore, 2012 and 2014). "High Nutrient" or "High Nitrate" conditions implied by the acronym HNLC then become exploitable in the upper few meters. Sea water flooding onto the Antarctic pack could support especially rich blooms, which may be more intense than those simulated here. Light appropriation by infiltrating organisms has not been obvious in our boreal simulations, but may apply across the Southern Ocean ice domain, where gap layers are deeply colored (Ackley and Sullivan, 1994; Thomas, 2003; Ackley et al. 2008). Since our mechanism includes metal cycling internal to the brine, we are prepared for rapid implementation to the south. Iron binding has been attributed to several subclasses of the biopolymers discussed here (Hassler and Schoemann, 2009; Van der Merwe et al. 2009) and they also have the ability to adsorb to walls of the ice network (Walstra, 2003; Janech et al. 2006; Underwood et al. 2010; McNeill et al. 2012). Detailed organometallic interactions will therefore constitute a new theme, in global simulations to be conducted by our group.

**Acknowledgements**: The authors thank the U.S. Department of Energy Office of Biological and Environmental Research (OBER) and specifically its ACME project (Accelerated Climate Model for Energy) for supporting the sea ice algal-chemical mechanism developments described here. Validation sections were funded in parallel by the OBER global biogeochemical program Benchmarking and Feedbacks, a department Science Focus Area (SFA).

**Table 1.** Arctic sea ice chlorophyll and dissolved organic carbon data, selected to guide and validate the reduced model presented here. The biogeographic breakdown in the left hand column follows Figure 1. Measured chlorophyll values are given in italics and units of mg/m³, while DOC is presented in bold as micromolar carbon. Within each cell the top, middle and bottom rows represent upper ice habitats, the interior and the bottom layer. Abbreviations: GIN –Greenland Iceland Norway (the collective Seas), NR –not reported, Beau –Beaufort Sea but mainly coastal, Arch –Canadian Archipelagic biogeographic province. Reference abbreviations: A01 (Amon et al. 2001), G97 (Gosselin et al. 1997), G99 (Gradinger, 1999), G09 (Gradinger, 2009), L05 (Lavoie et al. 2005), R97 (Robineau et al. 1997), R01 (Rysgaard et al. 2001), S88 (Smith et al. 1988), S97 (Smith et al. 1997), T95 (Thomas et al. 1995), U03 (Uzuka, 2003), WB89 (Welch and Bergmann, 1989). Comments: Where bottom layer data are reported in the literature as mg/m², a 3 centimeter thickness is adopted for conversion.

|  | Jan | Feb | Mar | Apr | May | Jun | Jul | Aug | Sep |
|---|---|---|---|---|---|---|---|---|---|
| **Okhotsk** |  | *300*, **100** | *1000*, **500** |  |  |  |  |  |  |
| **Bering** |  |  |  |  |  |  |  |  |  |
| **GIN** |  |  | *0*, **0** <br> *0.3*, **10** <br> *1*, **300** |  |  | **10** |  |  | *NR*, **100** |
| **Barents** |  |  |  |  |  |  |  |  |  |
| **Kara** |  |  |  |  |  |  |  |  |  |
| **Siberian** |  |  |  |  |  |  |  |  |  |
| **Chukchi** |  |  | *100* | *500* | *30* |  | *30* |  |  |
| **Beau** |  |  | *100* | *500* | *300* |  |  |  |  |
| **Arch** |  |  | *3* | *1000*, **300** | *2000*, **3000** | *300* |  |  |  |
| **Central** |  |  |  |  |  |  | *30* | *100* | *0* <br> *0.3* <br> *3* |
| Sources |  | S97 | R97, S97, T95, U03 WB89 | S97, U03 | G09, L05, S88, S97, U03 WB89 | L05, R01 | G97 | G97 | A01, G99 |

**Table 2**. Comparison of bottom layer chlorophyll maxima generated by a baseline model run, with the observations of Table 1 (mg/m$^3$). A fixed thickness of 3 centimeters applies in the case of the model. Geography, abbreviations and references as in the previous table; NB means no bloom. Blanks indicate the irrelevance of timing in the absence of growth or else a lack of data.

|  | Model | | Data | |
|---|---|---|---|---|
|  | Maximum | Month | Maximum | Month |
| **Okhotsk** | *600* | April | *1000* | March |
| **Bering** | *1100* | May | | |
| **GIN** | *NB* | | *10* | June |
| **Barents** | *500* | June | | |
| **Kara** | *800* | June | | |
| **Siberian** | *60* | June | | |
| **Chukchi** | *NB* | | *500* | April |
| **Beau** | *1200* | May | *500* | April |
| **Arch** | *900* | June | *2000* | May |
| **Central** | *NB* | | *100* | August |

**Table 3**. A summary of sea ice composition studies for bulk organic carbon contained in the brine. As might be expected, results cluster around established research stations. For many of the Figure 1 Arctic ecozones, data are thus quite sparse. A compilation of analogous Antarctic values is included toward the bottom of the table for comparison purposes. Polymers, oligomers and the associated monomers are considered together for the proteins and carbohydrates, and lipid data have so far been elusive. Beyond the first or dissolved column, results are agnostic toward phase state. All units are micromolar carbon. Abbreviations: DOC –dissolved organic carbon, Sacch –Polysaccharide or fresh carbohydrate, Arch and GIN as in earlier tables. Reference abbreviations: A01 (Amon et al. 2001), C01 (Calace et al. 2001), G00 (Guglielmo et al. 2000), H01 (Herborg et al. 2001), K02 (Krembs et al. 2002), N11 (Norman et al. 2011), R08 (Reidel et al. 2008), T01 (Thomas et al. 2001), Underwood et al. (2010). In this table, blanks carry forward information from directly above.

| Month | Zone | Level | DOC | Protein | Sacch | Lipid | Humic | Source |
|---|---|---|---|---|---|---|---|---|
| **March** | **Chukchi** | **Top** <br> **Interior** <br> **Bottom** | | | 50 <br> 100 <br> 50 | | | K02 |
| **April** | **Arch** | **Bottom** | 500 | | 100 | | | R08 |
| **May** | | | 1000 | | 500 | | | |
| | **Chukchi** | **Top** <br> **Interior** <br> **Bottom** | | | 250 <br> 250 <br> 250 | | | K02 |
| **September** | **GIN** | **Interior** | 100 | 5 | 15 | | | A01 |
| **Spring & Summer** | **Antarctic** | **Top** <br> **Interior** <br> **Bottom** | 100-300 <br> 50-300 | 50 | 20 <br> 50-100 | | 10 <br> 10 <br> 100 | C01, G00, H01, N11, T01, U10 |

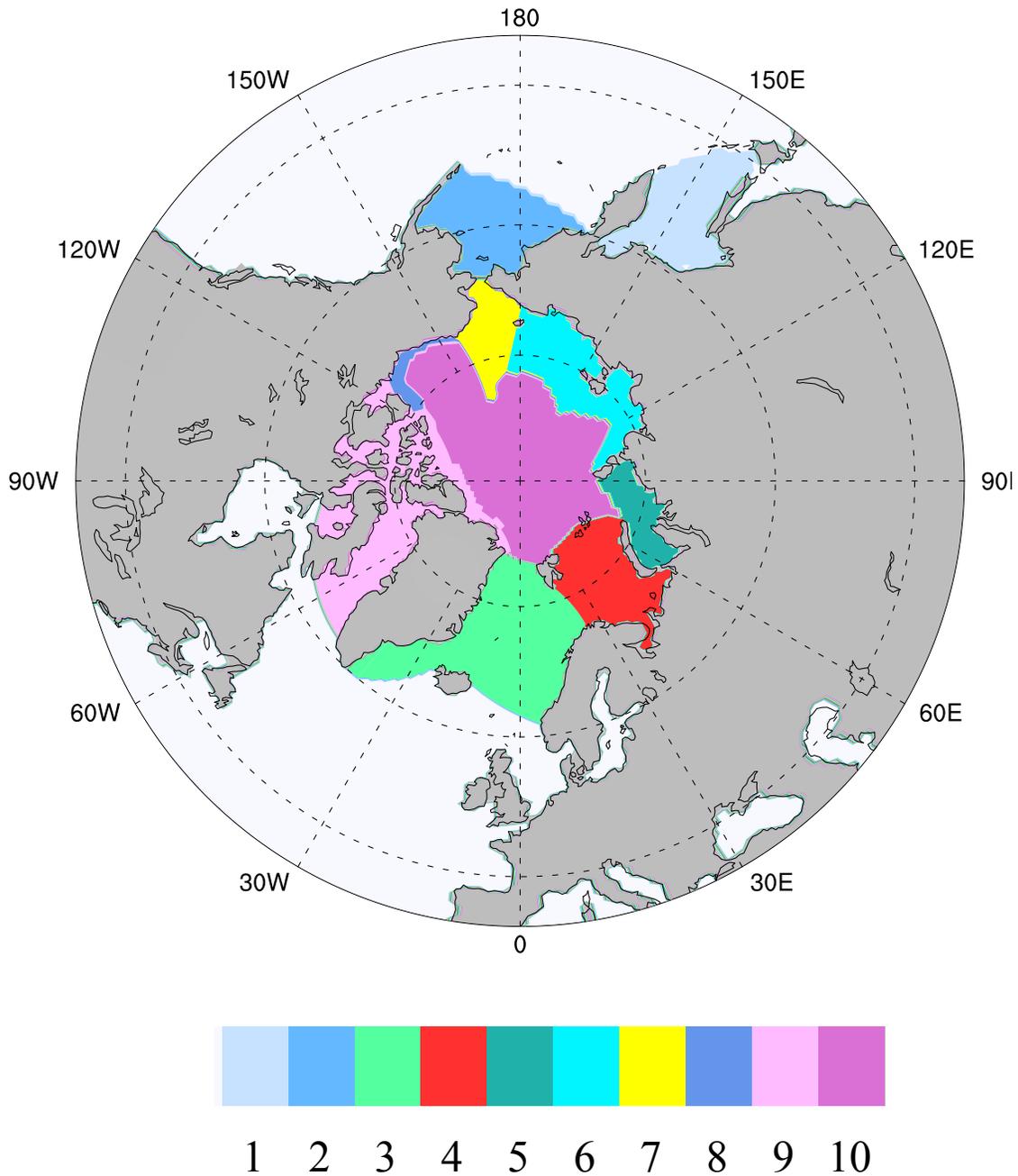

**Figure 1**. Ecogeographic zones over which an averaging of the CICE results was conducted, and within which the reduced scale ice algal modeling takes place. 1 – Sea of Okhotsk, 2 – the Bering Sea, 3 – GIN Seas (Greenland, Iceland, Norway), 4 – Barents Sea, 5 – Kara Sea, 6 – Siberian Shelf (combines Laptev and East Siberian), 7 – Chukchi Sea, 8 – Beaufort Shelf, 9 – Canadian Archipelago, 10 – Central Arctic to the pole. This system is an adaptation of Carmack and Wassmann (2006) as extended by Deal et al. (2011).

**Figure 2**. Schematic of an ice biogeochemical mechanism explored in the present work. Major ecodynamic channels are described anecdotally in the main text and as equations in the appendix area. Abbreviations: Diat –generic diatoms, Flag –microflagellates, Pha –*Phaeocystis* species, Chl –chlorophyll content for each of the previous algal classes, Prot –Proteins, Poly –the carbohydrates or polysaccharides, Lip –the broad molecular class of lipids, Hum –the heterogeneous polymeric mass of humic acid, Coll –colloids which may form from the organics, Ligs –Ligand effects which can become important under trace metal limitation, PAR – Photosynthetically Available Radiation.

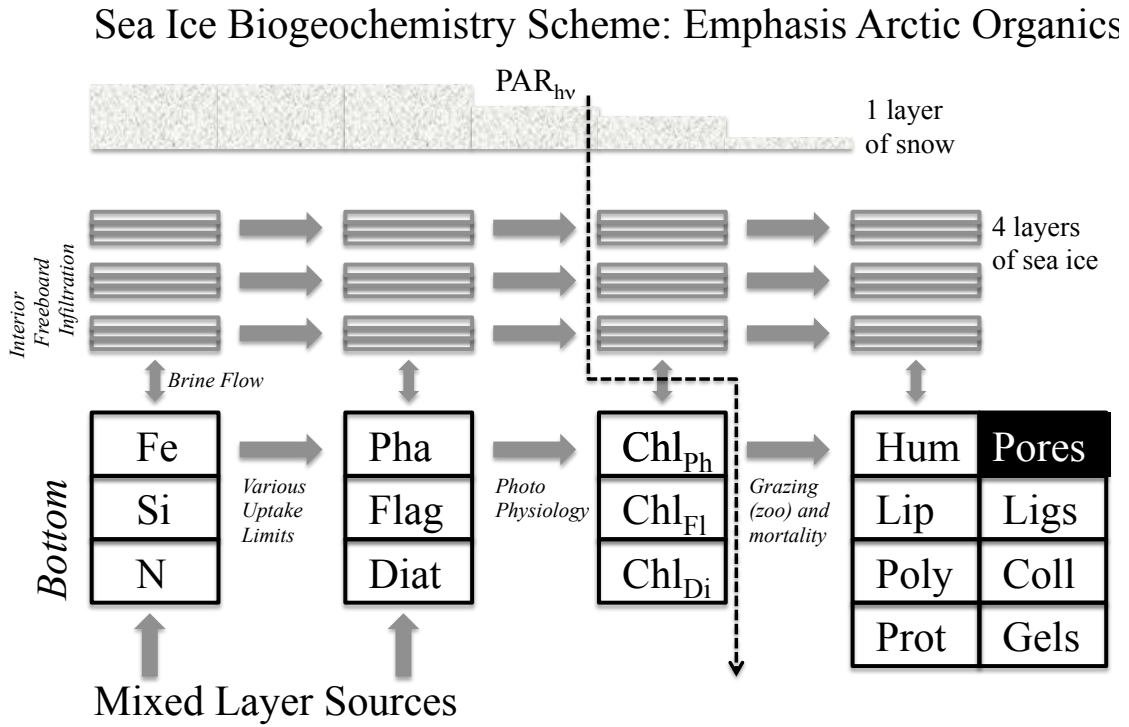

**Figure 3**. Sensitivity tests conducted in the reduced model with results presented in a cumulative format. To produce Panel A (left), bulk pigment concentrations (mg/m$^3$) were log transformed (base 10) and then Mean Absolute Error was summed along the ecogeographic list relative to Table 1 data. Average snow heights are superimposed for the springtime (meters x 10, linear). Jumps tend to occur where there is greater average precipitation. This is because bloom activity is shifted in certain regional cases. Upper habitat data are so rare that we calculated integrated chlorophyll-days for the infiltration layer as opposed to mean error (Panel B at right). A dearth of observations may indicate that the true value is near zero, or alternatively that the phenomenon is poorly documented in the Arctic. Thickness ratio used as a threshold was adjusted to vary the degree of flooding. Within the uncertainties entailed in snow height and density, infiltration can be largely eliminated.

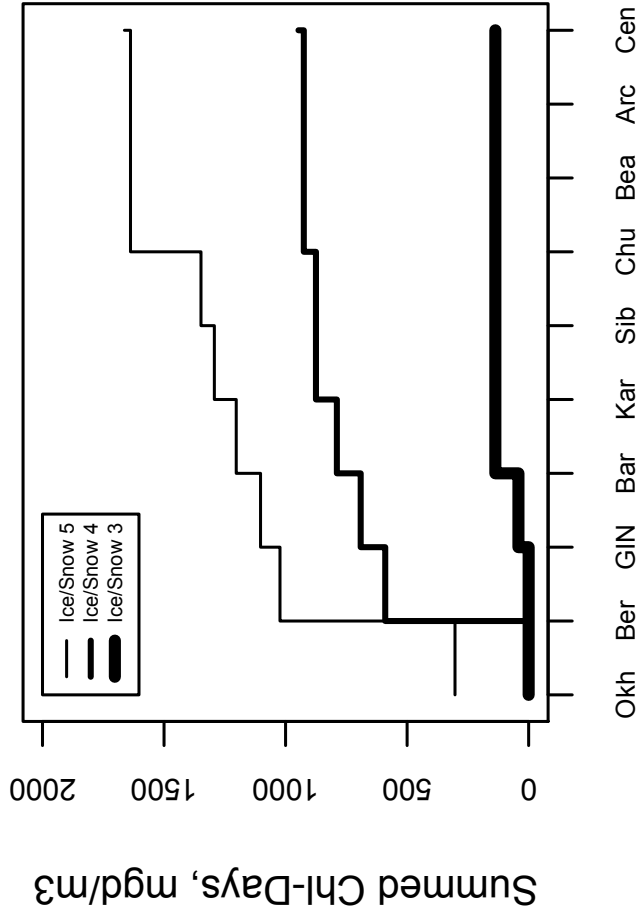

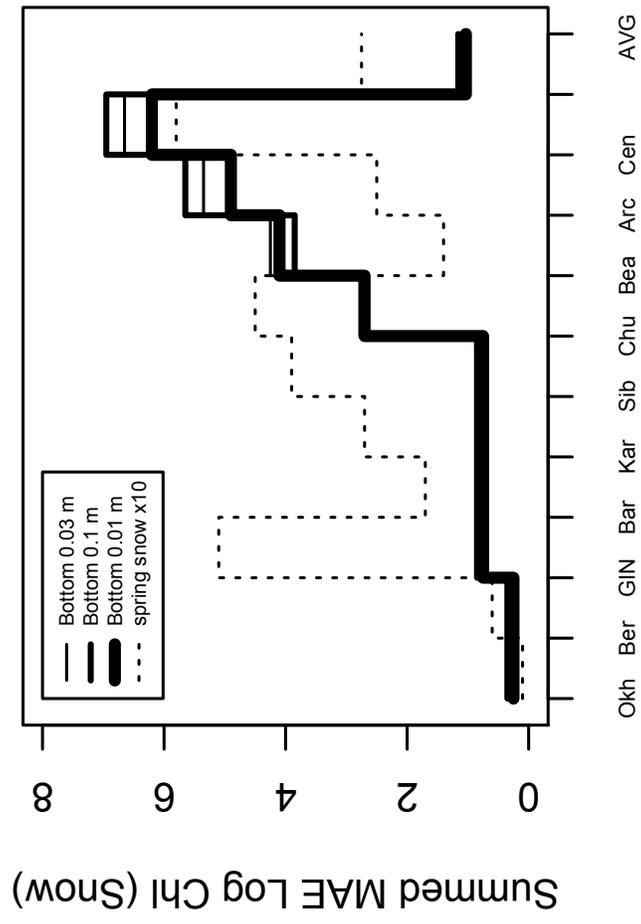

**Figure 4**. Sea of Okhotsk, time evolution for selected bulk nitrogen and carbon concentrations in the nutrient and ecodynamic tracer sets as output during the baseline run. All four numerical habitat levels are shown. All values are base 10 logarithms for the major indicator atom types and are derived from concentrations initially carried in micromolar, which is the model reference level (Appendix A). Silicate is not given because it was mainly present in excess. *Phaeocystis* also goes unrepresented, primarily in the interest of simplification. Its populations were close to those of the flagellates since there was little differentiation in their parameter settings (Appendix B).

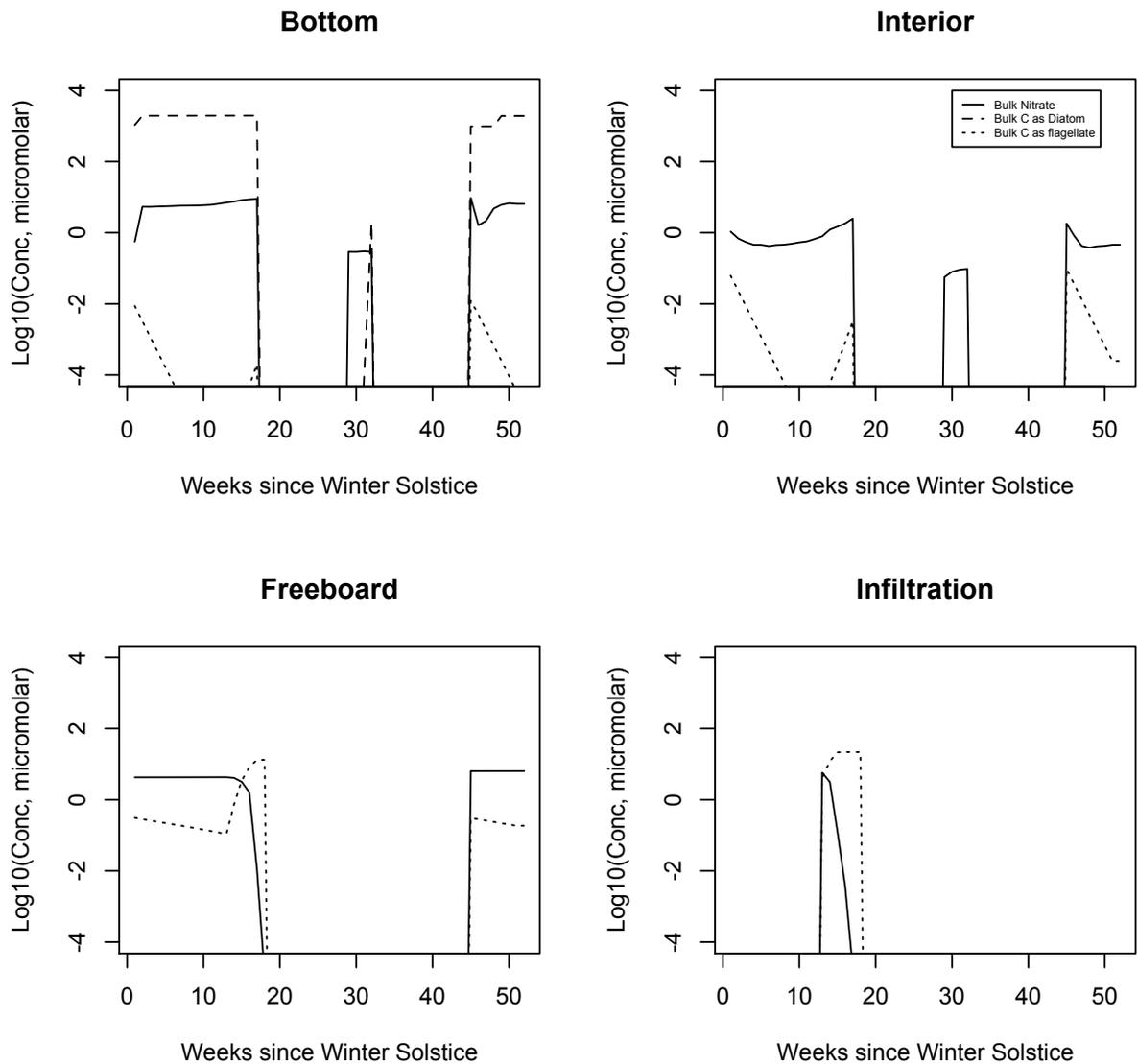

**Figure 5.** Chukchi Sea, caption as in 4.

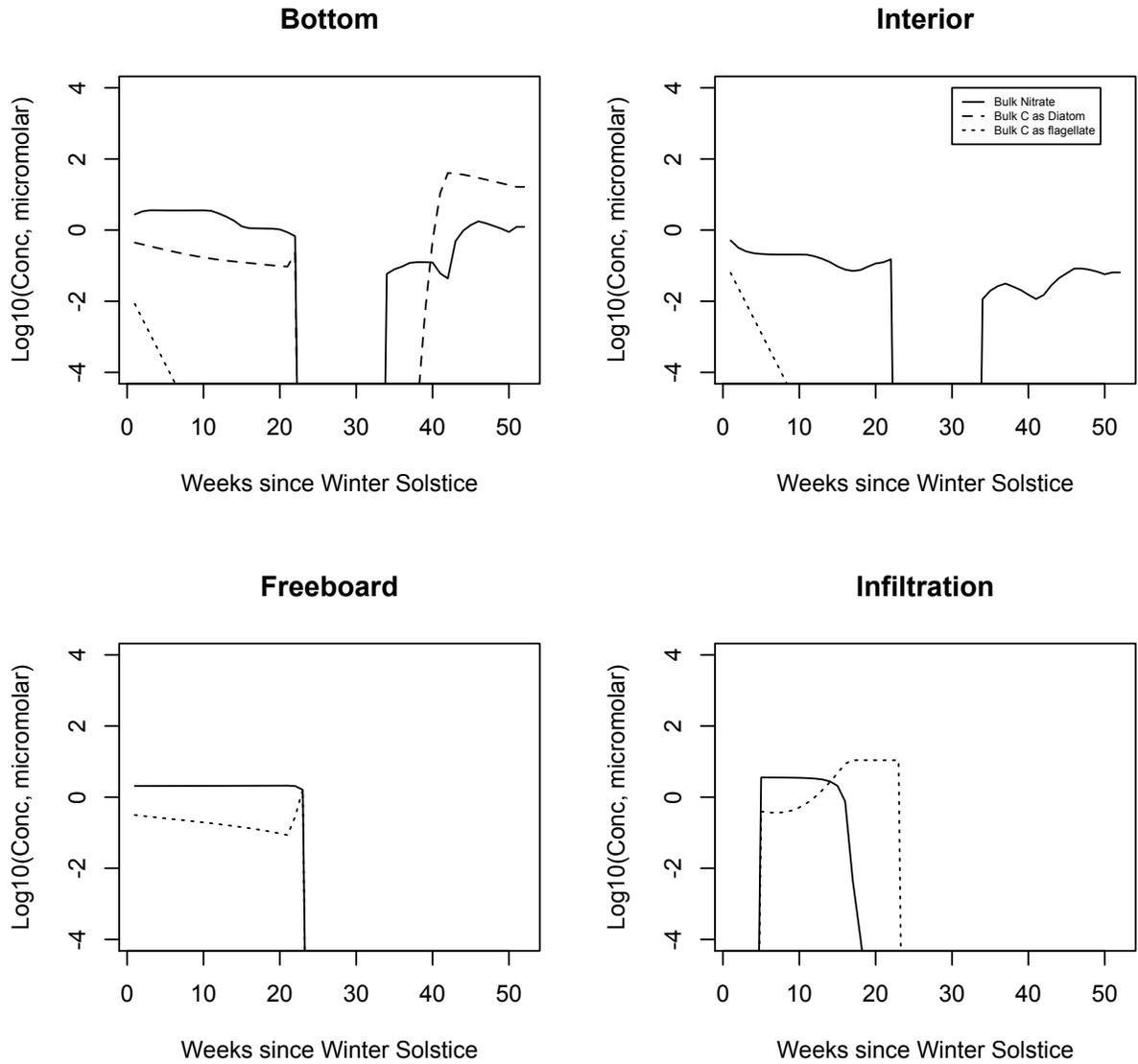

**Figure 6.** Beaufort Sea, caption as in 4.

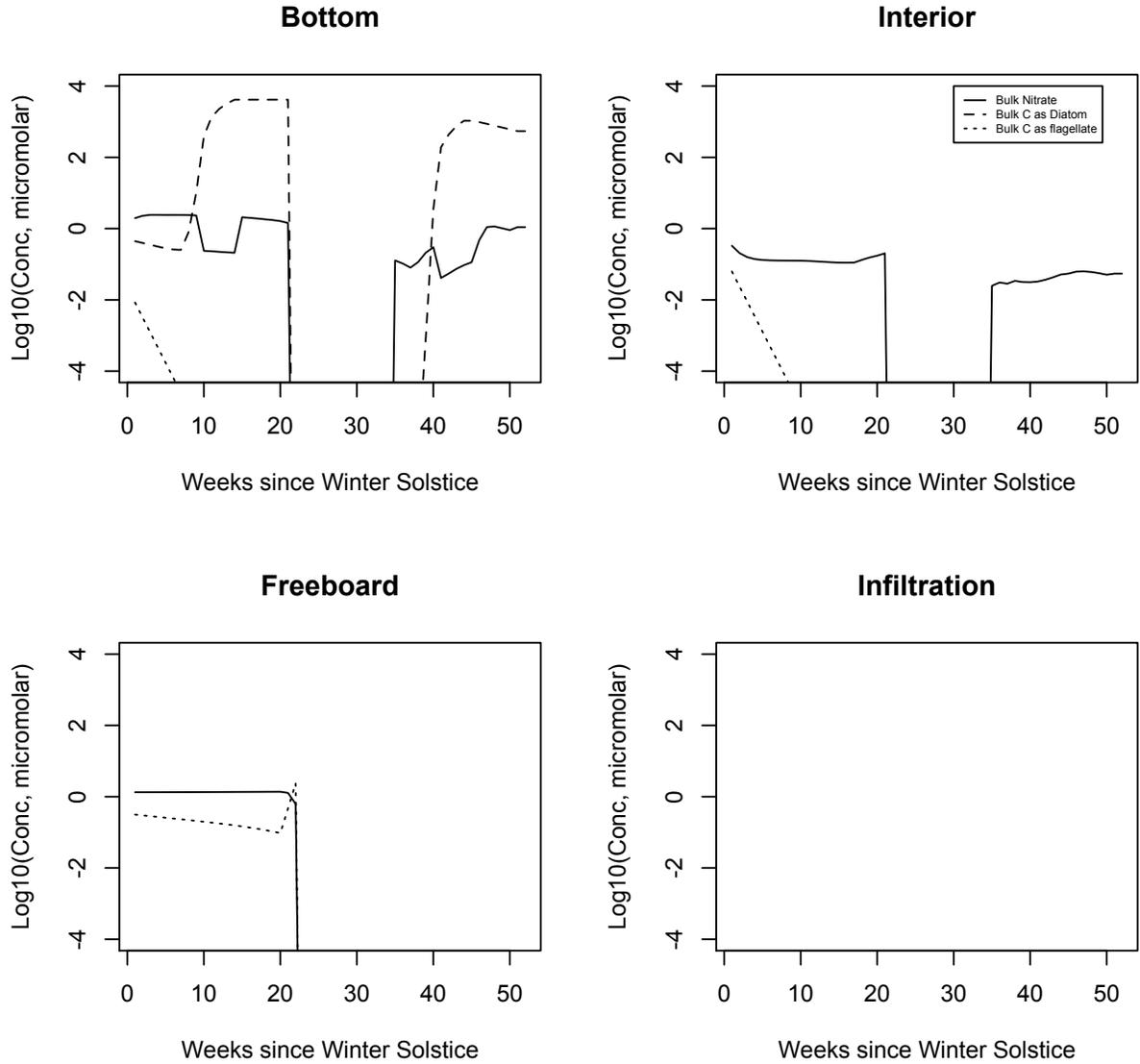

**Figure 7.** The pole, caption as in 4.

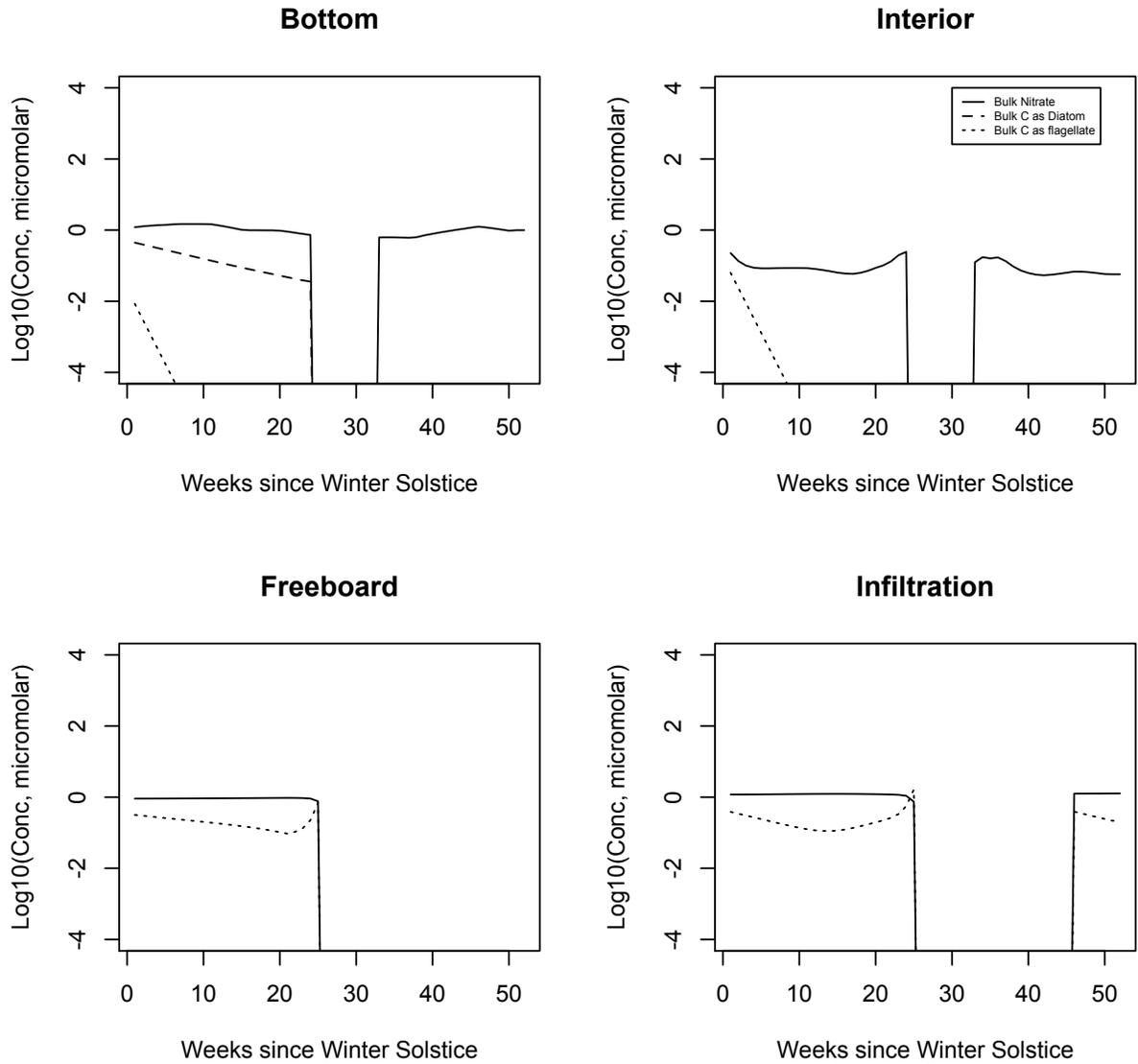

**Figure 8.** Sea of Okhotsk, time evolution for selected bulk compound concentrations from the organic tracer set as output during the baseline run. All four numerical habitat levels are shown. Values are base 10 logarithms for carbon and are derived from concentrations initially carried in micromolar, which is the model reference level (Appendix A). Lipid profiles closely follow those of the proteins or polysaccharides and so they have been deleted for simplicity.

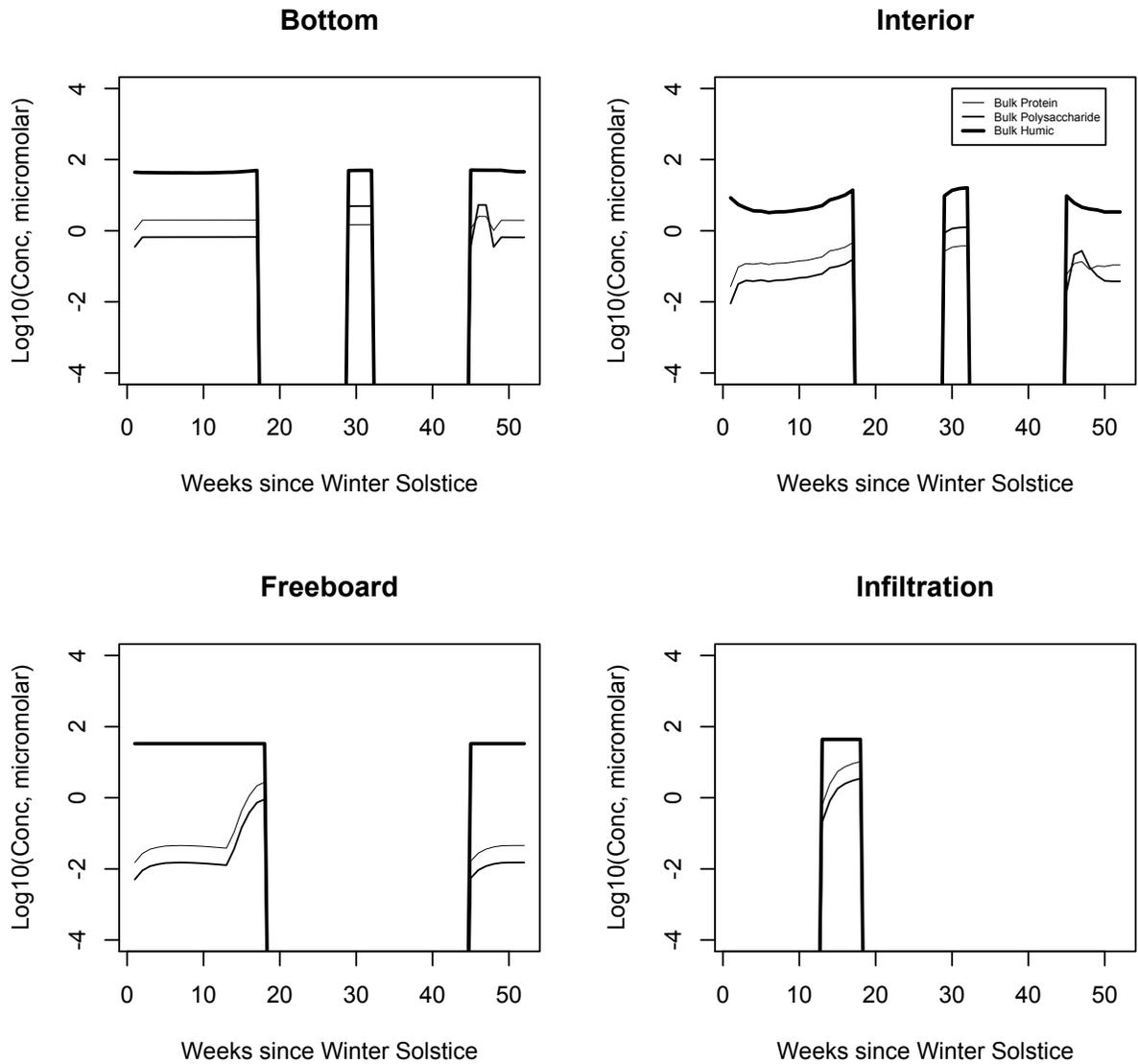

**Figure 9.** Chukchi Sea, caption as in 8.

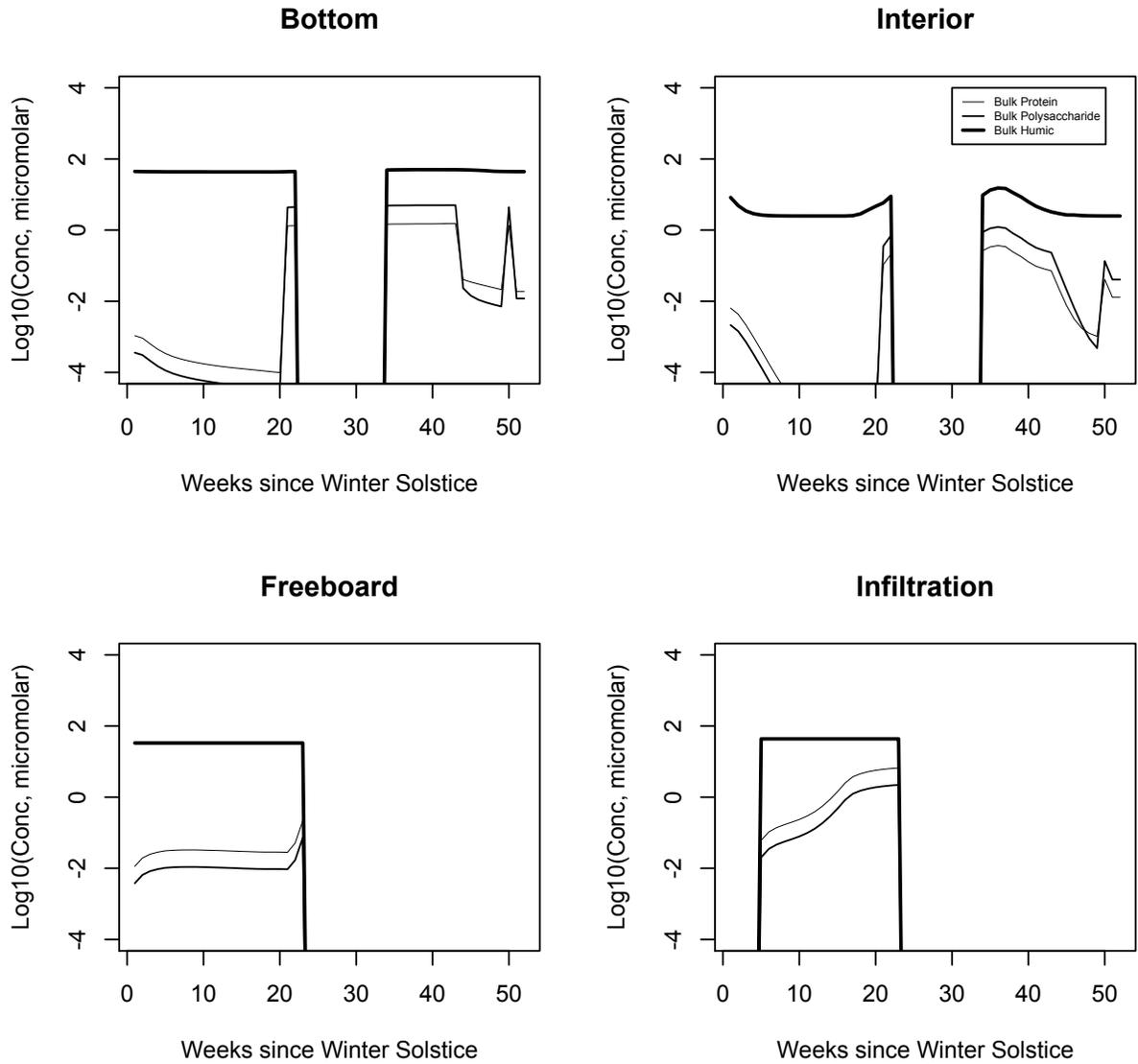

**Figure 10.** Sea of Okhotsk, time evolution for selected bulk compound concentrations from the organic tracer set as output during the light-dependent exudation run. All four numerical habitat levels are shown. Values are base 10 logarithms for carbon and are derived from concentrations initially carried in micromolar, which is the model reference level (Appendix A). Lipid profiles closely follow those of the polysaccharides and so they have been deleted for simplicity.

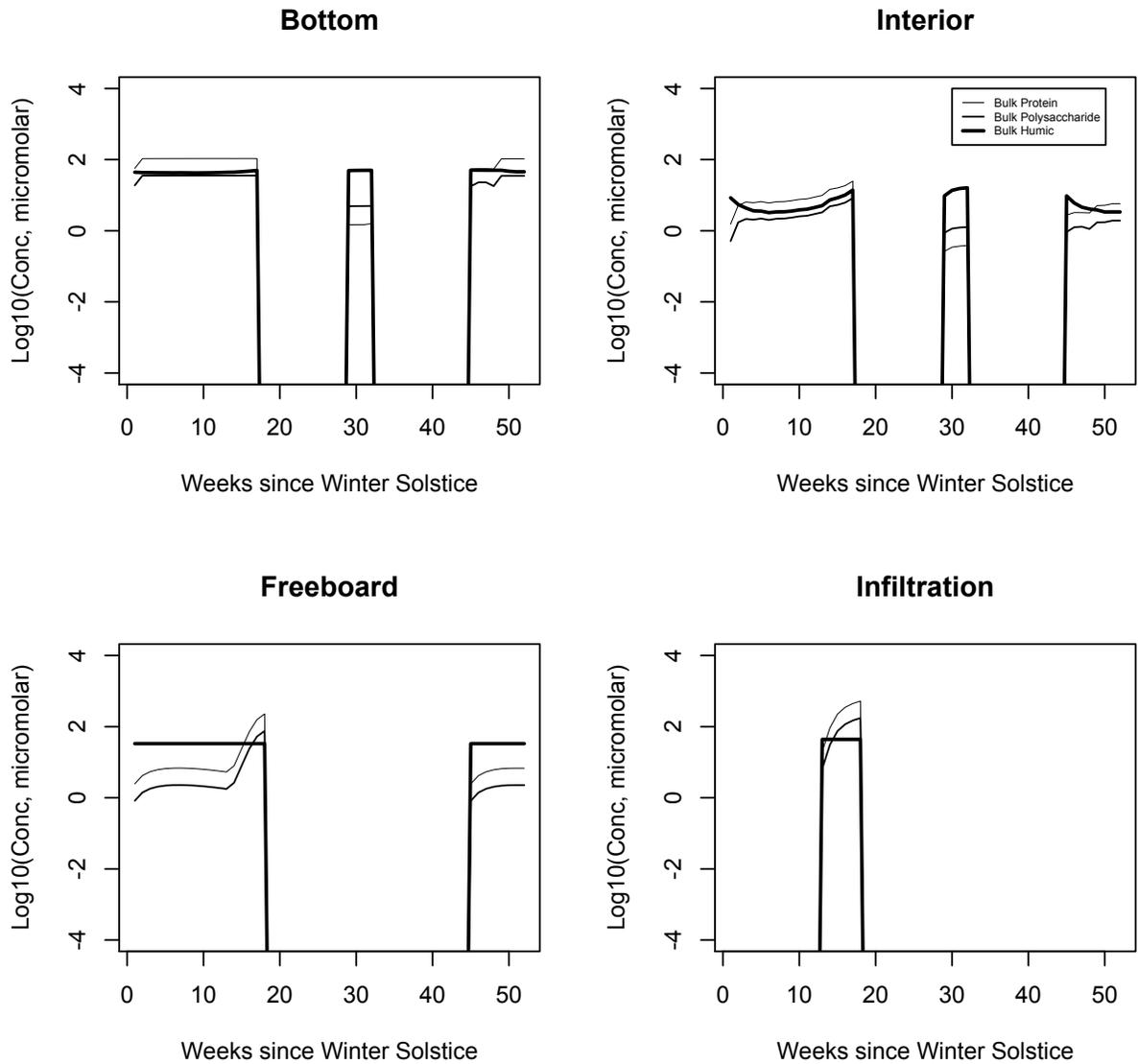

**Figure 11**. Chukchi Sea, caption as in 10.

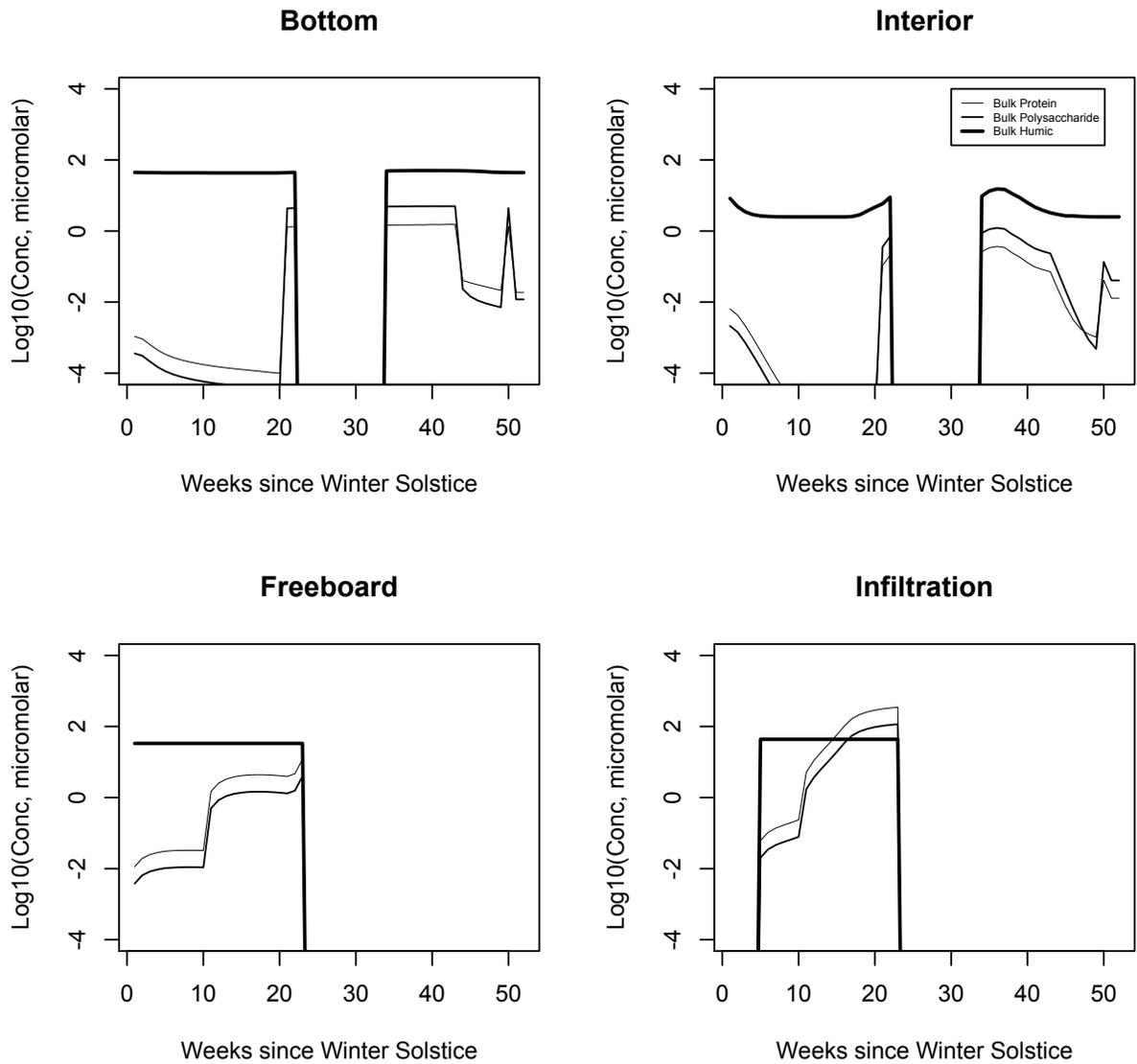

**Figure 12**. Beaufort Sea, caption as in 10.

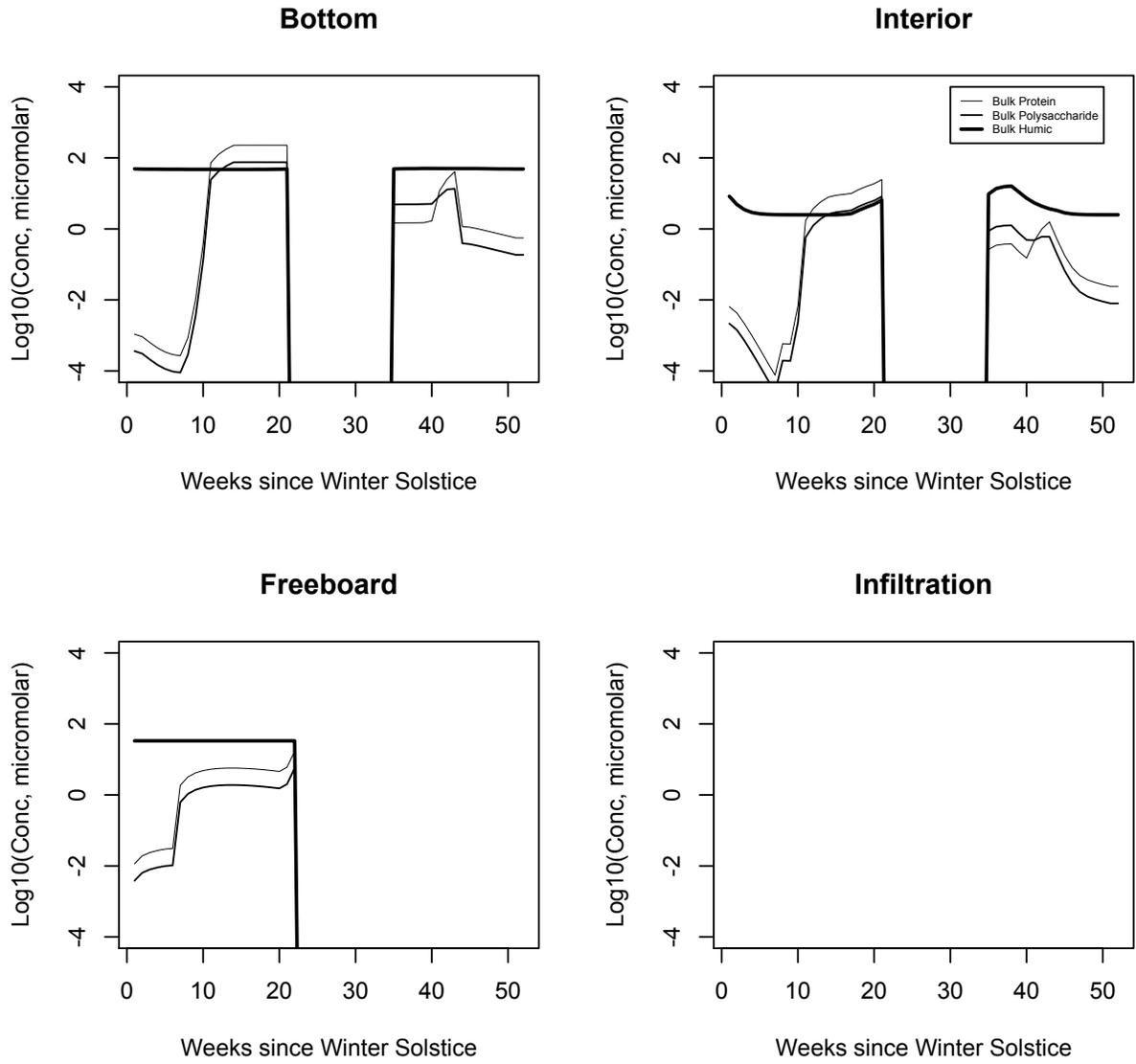

**Figure 13**. Pole, caption as in 10.

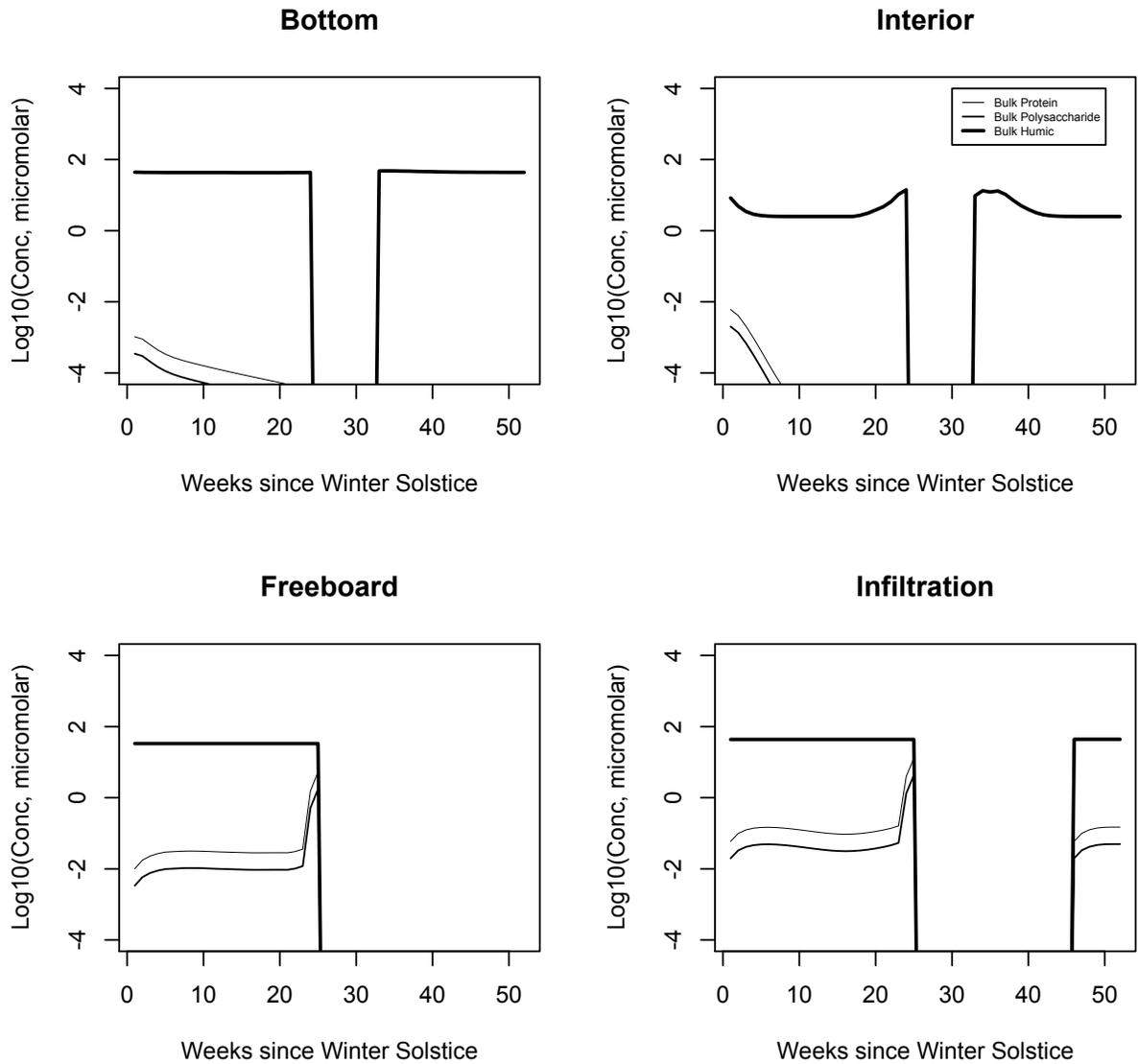

**Appendix A: Equations**

Concepts from the main text can be developed into equations representing the full evolution of ice internal ecodynamics along with dissolved organic chemistry. Although emphasis is placed upon Arctic environments in the present work, the ultimate intent is applicability for both polar regimes. Hence iron is included and considerable effort is devoted to upper level habitats. Until otherwise stated, all expressions are local to the interior of brine channels, so that nonparametric quantities are typically strong functions of vertical location $z$. For example nutrient nitrate $NO_3^- = NO_3^-(z)$, the internal light intensity $I_{avg} = I(z)$ and growth limitations $L^{type} = L(z)$. Constants in the system are provided in the parameter section. Essentially we combine the approaches of Fritsen et al. (1998), Lavoie et al. (2005), Jin et al. (2006) and Elliott et al. (2012), while acknowledging the pioneering work of Arrigo et al. (e.g. 1993 and 1997). Additional concepts are borrowed from the pelagic ecodynamics of Sarmiento et al. (1993), Walsh et al. (2001 and 2004), Gregg et al. (2003) and Moore et al. (2004).

In order to manage biogeochemical complexity, notational strategies are hybridized from matrix algebra, set theory and computer science. Bold font is reserved for ordered lists of related quantities, and these may be thought of as algebraic (nondirectional) vectors, tuples or computational arrays. Sub-superscript pairs are given either an organism-to-property/process relationship or else a kinetic to-from significance. All concentrations except chlorophyll are computed as millimole/m$^3$, which is conveniently identical to micromolar. This is always with reference to an element that is central to a molecule, unit of biomass or polymer chain. In early marine systems simulations, it was possible even at the ocean basin level to focus on a single atom as the primary currency (Sarmiento et al. 1993; Walsh et al. 2001). In this tradition we adopt carbon as a conventional choice to begin. But our purview extends to both multi-nutrient uptake and macromolecules of varied functional composition. The latter exhibit internal self-affinities as colloids or gels, plus a tendency to bind trace metals (Chin et al. 1998; Wells, 2002). Surface, intra- and intermolecular interactions are usually mediated by heteroatomic groups. Elemental ratios will thus be relevant, whether with regard to biological material or its detritus. Values are indicated by the convention $R^{N/C}$ etc.

Primary producers serve as a suitable initial example. It is anticipated that ultimately, bottom layer pennate diatoms, smaller autotrophs inhabiting central ice and even specialists such as *Phaeocystis* will be numerically segregated (Gradinger, 1999 and 2005; Fritsen et al. 2001; Lizotte, 2001; Tison et al. 2010; Underwood et al. 2010). Hence the subscript *i,auto* denotes the *ith* autotroph and our notation can be introduced as

$$\boldsymbol{auto} = pennate, smaller\ internal, Phaeocystis, other\ producers \tag{A1}$$

$$\boldsymbol{C}_{auto} = C_{pen}, C_{int}, C_{pha}, C_{other};\ R^{N/C}_{i,auto} C_{i,auto} = N_{i,auto},\ \boldsymbol{R}^{N/C}_{auto} \boldsymbol{C}_{auto} = \boldsymbol{N}_{auto} \tag{A2}$$

where *C* indicates net local carbon content of the brine. For the organisms themselves the ratios *R* may be thought of as classic Redfield relationships, but sometimes they will be permitted to vary from standard values (Walsh et al. 2001; Schoemann et al. 2005; Tedesco et al. 2012). Among the macromolecules, the elemental weightings should produce a Redfield average inside a cell (Parsons et al. 1984). But mixing and degradation in the aqueous medium lead rapidly to divergences from the norm (Thomas et al. 2001; Benner, 2002; Letscher et al. 2015). As usual, chlorophyll is tracked by its weight ratio to the content of the algae, as in $R^{chl/C}$ (Ackley et al. 1979; Fasham et al. 1993; Moore et al. 2002).

Mechanistic details extend well beyond the primary producers, so that multiple lists of related biogeochemical quantities are required. For example, we organize the nutrients and their central atomic constituents into arrays. This arrangement actually suggests that our informal vectors might be combined into matrices to some advantage. A list of nitrogen containing nutrients, for example, might consist of nitrate, ammonium, nitrite, amino acids and other forms. A broader strategy could potentially lead to automation of all the marine biogeochemistry, with compositions for the macromolecules arrayed as stoichiometries across the periodic table. For the moment however, there is a need to reserve the index $j$ for use in overall continuity expressions. In the few instances where a second dimension enters our reasoning, we will simply fix the atom type artificially.

Inorganics will be represented by a pseudo-elemental quantity $X$ excluding carbon and bearing the concentration of the central carrier. The dissolved organics by contrast must be simulated as polymers or in some cases chain aliphatics (Benner, 2002; Wells, 2002; Elliott et al. 2014). Therefore they are represented by their total carbon content as constituents of the brine. Our scheme maintains the ability to ratio against nitrogen, oxygen or sulfur as a means of counting functional groups (amine, hydroxyl, sulfide etc.). In the next few examples, *i,nut* signifies one of the several inorganic nutrients and *i,mac* a particular class of biomacromolecule. Iron is not speciated at this point. In fact its concentrations are set high to discourage trace metal limitation, which is rare in the Arctic. Ultimately a goal is to resolve chelation chemistry per Tagliabue et al. (2009), Hassler and Schoemann (2009), Hassler et al. (2011) and other investigations taking ligands into account.

$$\boldsymbol{nut} = NO_3^-, NH_4^+, Si(OH)_{4,} Fe^{avail}; \boldsymbol{ele} = N, Si, Fe \ (ignore \ O, H) \tag{A3}$$

$$\boldsymbol{X}_{nut} = X_{nit}, X_{amm}, X_{sil}, X_{fe}; \ X_{1,nut} = NO_3^- \tag{A4}$$

$$\boldsymbol{mac} = protein, saccharide, humic, siderophore, etc.; \ C_{1,mac} = C_{prot} \tag{A5}$$

Using this array notation, we now address general ice ecodynamics working from the top down, by attenuating incoming solar radiation. The approximation is made that when photons penetrate snow and interact optically that they have been absorbed. Scattering will be available in the full CICE model. Light limitation is computed avoiding photophysiological factors by noting that chlorophyll absorption depends only weakly on pigment packaging (Arrigo and Sullivan, 1992). Spectral resolution is averaged into a traditional broad band spanning the visible wavelengths and referred to as Photosynthetically Available Radiation or PAR (Parsons et al 1984; Lavoie et al. 2005; Elliott et al. 2012). In what follows $I_{avg}$ is the total intensity integrated across the visible portion of the spectrum, in units of W/m². The italics *s* and *in* refer to saturation and inhibition scaling. The saturation reference point is adjusted for acclimation moving downward away from light sources (Arrigo and Sullivan 1992; Arrigo et al. 1993) through a linearization of culture data. Since step sizes in the offline code approach or exceed environmental adaptation times, the adjustments can be continuous.

$$I_{avg} = (I_o/l_{path}) \int_o^{l_{path}} exp(-al)\, dl; a = a_w + \left(\sum_{i,auto} a_{i,auto}\, Chl_{i,auto}\right) \quad (A6)$$

$$L_{i,auto}^{rad} = (1 - \exp(-(\alpha_{i,auto}/P_{i,auto}^{max})I_{avg})\exp(-(\beta_{i,auto}/P_{i,auto}^{max})I_{avg}) \quad (A7)$$

$$(\alpha_{i,auto}/P_{i,auto}^{max})^{-1} = I_{i,auto}^s;\ (\beta_{i,auto}/P_{i,auto}^{max})^{-1} = I_{i,auto}^{in};\ I^s(PAR = I_{avg}) \quad (A8)$$

where $a_w$ is the total attenuation by either snow or ice, with units of reciprocal distance (Lavoie et al. 2005). Nutrient terms are analogous with the light limitation $L^{rad}$, but we draw on the $X$ to construct Monod factors across the inorganic concentration vector. Fractional apportionments ranging from 0 to 1 begin to appear and are generically represented as $f$. Here in the production calculations they refer mainly to the proportion of growth supported by a given nutrient type inside an elemental class (Elliott et al. 2012). Only nitrogen possesses two bioavailable forms in our expressions, so the central atom type is fixed in this case to distribute across oxidation states. Deeper in the growth equation, we include a reduction term for physiological effects of extreme salinity at low brine temperatures (Arrigo and Sullivan, 1992; Arrigo et al. 1993; Arrigo, 2003).

$$L_{i,auto}^{i,nut} = (X_{i,nut}/(X_{i,nut} + K_{i,auto}^{i,nut}))exp(-\chi_{i,nut}NH_4^+) \quad (A9)$$

$$L_{i,auto}^{i,ele} = \left(\sum_{i,nut} L_{i,auto}^{i,nut}\right)\Big|_{i,ele},\ (prorate\ to\ maximum\ of\ unity) \quad (A10)$$

$$f_{i,auto}^{NO_3^-,NH_4^+} = (L_{i,auto}^{i,nut}/L_{i,auto}^{i,ele})_{capped},\ f_{i,auto}^{Si(OH)_4} = f_{i,auto}^{Fe\,avail} = 1 \quad (A11)$$

$$L_{i,auto}^{total} = Min(L_{i,auto}^{rad}, L_{i,auto}^{ele}) \quad (A12)$$

$$Growth_{i,auto}^C = L_{i,auto}^{total} g_{i,auto}^{pre} exp(g_{i,auto}^{exp} Temp^o) f_{i,auto}^{sal} C_{i,auto} \quad (A13)$$

$$Growth_{i,auto}^N = R_{i,auto}^{N/C} Growth_{i,auto}^C;\ similarly\ for\ other\ elements, chlorophyll \quad (A14)$$

Following Jin et al. (2006) then Elliott et al. (2012), the factor $\chi$ is set to zero in the above excepting nitrate ($\chi_{nut}$ = 1.5 1/μM, 0, 0, 0). The concept is to simulate uptake inhibition by the more useful reduced nitrogen form (Fasham et al. 1993). In this latest set of equations, growth is a shorthand for gross primary production and has the units of a chemical rate (millimole/m³s or per day). In all cases temperature must be expressed in Celsius degrees (°).

We now proceed to the construction of ecosystem flow terms which rely heavily on fractionation concepts. The abbreviations "resp, mort, spill, zoo, assim, excr and remin" are shortened from the terms respiration, mortality, spillage, zooplankton, assimilation, excretion and remineralization respectively. For more detailed discussion of such routings see Fasham et al. (1993) or Moore et al. (2002 and 2004) then Elliott et al. (2012). One of the main goals of future generation ice and global ecodynamics simulation will be to render the *f* below more dynamic and explicit.

$$dC_{i,auto}/dt = \left(1 - f_{i,auto}^{graze} - f_{i,auto}^{resp}\right)Growth_{i,auto}^{C} - Mort_{i,auto}^{C} \tag{A15}$$

$$Mort_{i,auto}^{C} = m_{i,auto}^{pre} exp\left(m_{i,auto}^{exp} Temp^{o}\right)C_{i,auto} \tag{A16}$$

$$dX_{i,auto}^{ele}/dt = R_{i,auto}^{ele/C}\left(dC_{i,auto}/dt\right) \tag{A17}$$

In the equations immediately following, the identifier *ele* is carried implicitly for clarity, e.g. $Resp^{ele} = f^{resp} Growth^{ele}$.

$$Resp_{i,auto} = f_{i,auto}^{resp} Growth_{i,auto}; Graze_{i,auto} = f_{i,auto}^{graze} Growth_{i,auto} \tag{A18}$$

$$Spill_{i,auto} = f_{zoo}^{spill} Graze_{i,auto}; \quad Assim_{i,auto} = f_{zoo}^{assim} Graze_{i,auto} \qquad (A19)$$

$$Excr_{zoo} = f_{zoo}^{excr} Assim_{zoo}; \quad Assim_{zoo} = \sum_i Assim_{i,auto} \qquad (A20)$$

$$Remin_{i,auto} = f_{i,auto}^{remin} Mort_{i,auto} \qquad (A21)$$

$$f_{i,auto}^{resp} + f_{i,auto}^{graze} \leq 1; \quad f_{zoo}^{spill} + f_{zoo}^{assim} = 1; \quad f_{zoo}^{excr} \leq 1 \qquad (A22)$$

The factors $f$ ultimately become quite ad hoc, but this is a typical expedient in marine biogeochemistry models (Sarmiento et al. 1993; Arrigo et al. 1993; Christian and Anderson, 2002; Moore et al. 2002; Tedesco et al. 2012). The advantage is that sensitivity testing is rendered convenient and interpretable. We treat the zooplankton as a single entity so that no subscripting is required for classes or trophic levels. The consumer organisms actually go un-modeled. Since grazing has often been treated as a proportion of growth within sea ice, dynamics of the secondary producers are deferred (Arrigo et al. 2003; Lavoie et al. 2005; Jin et al. 2006; Elliott et al. 2012). See a localized study by Tedesco et al. (2012) for more comprehensive alternatives, discussed in relation to the ecodynamics of a Greenland fjord.

Establishment of autotrophic populations implies uptake of nutrients. The case of silicate is trivial because this solute tends to be consumed and never released from the solid (Lavoie et al. 2005 and 2009). See Elliott et al. (2012) for a discussion of the fate of organism hard parts. Mainly they consist of frustules of the ice bottom-dwelling pennate diatoms. The nitrogen inorganics are interrelated by recycling processes. Oxidation states are linked to one another along the overall metabolism then by microbial chemistry (Fritsen et al. 2001; Thomas and Papadimitriou, 2003), and routings are once again prominent in the $NH_4^+$ case. We model nitrification after Jin et al. (2006) and all of respiration is presumed to flow through ammonia.

No attempt is made to reserve nitrogen atoms for a conservative production of proteins. We view this as a potential application of upcoming atom-balancing approaches. Iron recycling is treated by analogy with ammonia for the simple reason that detailed ice chemical information is lacking (Van der Merwe et al. 2009; Lannuzel et al. 2010).

$$Uptake_{i,auto}^{i,nut} = R_{i,auto}^{ele/C} f_{i,auto}^{i,nut} Growth_{i,auto}^{C} \tag{A23}$$

$$dX_{i,nut}/dt = -\sum_{i,auto} Uptake_{i,auto}^{i,nut} - Loss_{i,nut} + Recycle_{i,nut} \tag{A24}$$

$$Loss_{nit} = 0;\ Loss_{am} = k_{nit} NH_4^+;\ Loss_{Fe} = 0 \tag{A25}$$

$$Recycle_{nit} = k_{nit} NH_4^+ \tag{A26}$$

$$Recycle_{am} = \sum_{i,auto} Resp_{i,auto}^{N} + \sum_{i,auto} Remin_{i,auto}^{N} + Excr_{zoo}^{N};\ Fe\ likewise \tag{A27}$$

We have now dealt with most ecodynamic pathways associated with primary and secondary production. Macromolecules in the dissolved form or else as colloidal, gelling and chelating agents are now added to the scheme. We begin with the list of compounds which has proven useful in studies of global surfactants at the air water interface (Benner, 2002; Elliott et al. 2014; Burrows et al. 2014). To this may be added specific intentional releases including polysaccharides (Passow et al. 1994), aminosugars (Van Rijssel et al. 2000), siderophores (Hassler et al. 2011) and more. Generic exudation terms are incorporated for sensitivity testing, with rate constants functionalized as in the text. Since we assume that either leaks are small or light energy may be harvested to support additional fixation, exuded carbon need not be accounted during mass conversation. Working from ***mac*** and ***C**mac* above

$$C_{mac} = \sum_{i,mac} C_{i,mac};\ \boldsymbol{R}_{mac}^{N/C} \boldsymbol{C}_{mac} = \boldsymbol{N}_{mac};\ similarly\ for\ O, P, S, metals \tag{A28}$$

$$Release_{i,mac}^{C} = f_{i,mac}^{biomass}(Spill_{auto}^{C} + Mort_{auto}^{C}) + Exude_{i,mac}^{C} \tag{A29}$$

$$Exude_{i,mac}^{C} = \sum_{i,auto} k_{i,auto}^{exu} f_{i,mac}^{biomass} C_{i,auto} \tag{A30}$$

$$Consume_{i,mac} = k_{i,mac}^{bac} C_{i,mac} \tag{A31}$$

$$dC_{i,mac}/dt = (Release - Consume)_{i.mac} \tag{A32}$$

The biomass fractionations $f_{i,mac}$ are arranged over $i$ to sum to unity and to reflect the standard carbon ratios inside algal cells, which are fairly uniform for the proteins, polysaccharides and lipids (Parsons et al. 1984; Wakeham et al. 1997; Benner, 2002). We assume that high-quality fresh macromolecules and polymers are not excreted by the zooplankton. Removal is mainly bacterial and rates are based on empirical data from real ice systems (Thomas et al. 1995; Smith et al. 1997; Amon et al. 2001; Reidel et al. 2008). Microbial heterotrophs sometimes decouple strongly from the organic cycling within ice at low temperatures (Pomeroy and Wiebe, 2001), and their behaviors are unfamiliar in other ways (Grossman and Gleitz, 1993). Incorporation during frazil stages may require physical attachment to symbiotic algae. Metabolisms may be strongly reduced during frazil/pancake formation, but as congelation proceeds the microbes recover quickly, establishing a growth constant of close to one day and exhibiting relatively low mortality (Grossmann and Dieckmann, 1994). We assume that bacteria act upon the organic reservoir in a unimolecular fashion. For example, Arrigo et al. (1995) discuss exponential vertical protein profiles while Amon et al. (2001) provide data consistent with a one month oxidative time scale in Arctic ice samples.

Local time tendencies can be prepared in this manner for all quantities, and so we turn attention to transport. Vertical motions are dealt with exclusively, since over the course of a bloom horizontal advection is slow (Nowlin and Klinck, 1986; Rampal et al. 2009). Outline font now distinguishes a bulk ice quantity from its counterpart in the brine (Vancoppenolle et al. 2010). A vertical continuity equation should apply to all our reactive tracers, and they will now be denoted $\mathbb{RT}$. Thus the biogeochemical tuple becomes ***RT = X, C*** if we think in terms of central atoms such as silicon or carbon. A bulk (outline) analog vector follows directly. Flux is given separately to emphasize the role of porosity in regulating net mixing. Our approach is based heavily on Vancoppenolle et al. (2010), but all critical quantities are available in CICE (Hunke et al. 2015) and they can be transported there as well (Jeffery et al. 2011; Elliott et al. 2012). The brine channel system is assumed to be isotropic.

$$\partial \mathbb{RT}_{i,bgc}/\partial t = - \partial \mathbb{F}_{i,bgc}/\partial z + \mathbb{P}_{i,bgc}(\mathbb{RT}_{j,bgc}, I_{avg}) - \mathbb{L}_{i,bgc}(\mathbb{RT}_{j,bgc}, I_{avg}) \tag{A32}$$

$$\mathbb{F}_{i,bgc} = f^{bri}\big(-K(\partial RT_{i,bgc}/\partial z) + v^{melt} RT_{i,bgc}\big);\ f^{bri} = V^{bri}\ as\ fraction \tag{A33}$$

$$(\mathbb{P} - \mathbb{L})_{i,bgc} = f^{bri}(dRT_{i,bgc}/dt);\ \mathbb{RT}_{i,bgc} = f^{bri} RT_{i,bgc} \tag{A34}$$

New symbols introduced include *bgc* identifying the summed biogeochemical tracer list, *F* for flux, *P* and *L* for production/loss, j representing all interactants, superscript *bri* for a brine specific quantity, *K* as a generic diffusion coefficient, a flushing velocity driven by upper level melting, and *V* for brine volume. The purpose for introducing outline font quantities is to combine brine channel kinetics with bulk mixing considerations. The *z* axis is defined to be positive upward so that $v^{melt}$ is negative. Production and loss terms computed as ordinary (time) differentials must be downgraded by the volume fraction for bulk computation.

At this point we have a representation of column chemistry-transport for global classes of nutrients, ice algae and detrital macromolecular carbon -all packaged as a vertical scheme for distribution through sea ice. A next challenge will be to develop the equations into a fully dynamic biogeochemistry model (Hunke et al. 2015). For early testing, a reduced form is now described which runs entirely offline. It is also portable to arbitrary output from other groups, so that intercomparison is facilitated. Estimated distributions for the ice algae and their byproducts can be constructed in a convenient post processing mode.

Certain terms can be handled quickly based on a typical sequence of events as spring unfolds. For example, $v$ is nearly zero until snow melt begins. Eventually a nearly instantaneous purge may dominate advection (Eicken et al. 2002; Jin et al. 2006). Vancoppenolle et al. (2010) model the total rate as

$$f^{bri} v^{melt} = \delta(Temp^o) f^{perc} R^\rho (dh/dt)|_{surf} \tag{A35}$$

$$v^{melt} = (Melt - Leads)\delta(Temp^o)/f^{bri} \tag{A36}$$

The delta function is set to unity if the rule of 5's allows communication between channels (Golden et al. 1998). Here $R$ is the ratio of densities for snow conversion, $h$ is a height and finally $f^{perc}$ is the fraction percolating into the network or in other words, not sloughed into cracks or leads (Eicken et al. 2002). The $v^{melt}$ may be considered sudden, so that ice internal biogeochemical activity shuts down when/where significant surface loss and the delta phase transition coincide. If the time scale $h/v^{melt}$ is less than one week we remove all concentrations. The flux equation therefore simplifies to a single diffusion term.

A minimum of four layers is needed to pick up major habitat types: bottom, interior, freeboard and infiltration zones (Ackley and Sullivan, 1994; Melnikov, 1997; Haas et al. 2001; Arrigo, 2003). The latter two may be difficult to separate observationally since they are proximate and measurements tend to be made during spring when they coincide and are decomposing (Ackley and Sullivan, 1994). However, the freeboard and infiltration processes are distinct so that we treat them independently. In our transport framework, the outer partial in depth becomes a thickness while its inner counterpart now represents a laminar layer boundary. The dimension of all layers other than the interior is initialized at 3 centimeters based primarily on Arctic studies (Reeburgh, 1984; Arrigo et al. 1997; Melnikov, 1997; Lavoie et al. 2005; Elliott et al. 2012). In our reduced framework, the default interior extent is 30 centimeters (Gradinger, 1999 and 2005). The bottom and top of the column tend to be active because they are close to nutrient and light sources respectively, but interior habitats cannot be ignored (Arrigo, 2003; Gradinger et al. 2005). Ice growth physics is intricate for all the layers (Cox and Weeks, 1975; Ackley and Sullivan, 1994; Lavoie et al. 2005), but in our approach to the biogeochemical kinetics dimensions are kept fixed during each simulation. The bottom layer remains in constant communication with the ocean except during the melt (Jin et al. 2006; Elliott et al. 2012). The content of the interior mixes with bottom ice but reactions take place independently. Freeboard material is positioned above the porosity cut off, and so blooms typically form under stringent initial mass limitations (Ackley and Sullivan, 1994). A combination of upward percolation and flooding saturates the top of the ice when it is weighted down sufficiently by snow (infiltration -Fritsen et al. 1998). The threshold for submersion is determined by Archimedes Principle, and a possibility of multiple flooding events was explored. Snow cover, ice thickness and temperature profiles are all extracted as time averages from dynamic model output (Hunke et al. 2015). By contrast with the chemistry, radiation penetration takes CICE depth variation into account. The resolution of ingested history files was one week.

The barrier to vertical transport is taken in all cases to be a laminar layer segregating boxes well mixed within themselves due to brine convection. For the ocean-to-bottom reservoir transition (*o2b*) we mimic Lavoie et al. (2005) and place resistance in the sea, just below the (baseline) three centimeter thickness of porous material. The laminar dimension is adjusted to a transfer velocity $f^{bri}K/\Delta z^{o2b} = v^{o2b}$ of 0.3 m/d as a starting point. Geographically, interchange will vary with details of boundary layer physics, including shear and tidal effects. Crossing from the bottom layer to the interior, tracers must pass through a zone of rapidly falling porosity. The relevant geometry is that of established brine channels, typically one tenth millimeter across (Perovich and Gow, 1991; Light et al. 2003; Krembs et al. 2011). Molecular diffusion will define the minimum rate of passage at about $10^{-9}$ m$^2$/s for most solutes (Stumm and Morgan, 1981; Lavoie et al. 2005). We treat the vertical extent of the resistance zone as a variable parameter with a starting round figure value of one millimeter. Thus the calculations begin at $f^{bri}K/\Delta z^{b2i} = v^{b2i} = f^{bri}0.1$ m/d. Porosity is averaged from internal CICE data (Jeffery et al. 2011; Hunke et al. 2015) so that transfer is initialized at a level significantly slower than for entry into the bottom layer, viewed from the bulk perspective. Roughly speaking a meter of typical ice thickness fills with biogeochemical activity in order weeks. Upper layers are treated as isolated kinetic boxes because they cap solid material which is sometimes closed due to the rule of fives. The laminar barrier may therefore be conceived of as infinitely thick. This is consistent with the observation that surface chlorophyll bands may be trapped and very intense (Ackley et al. 2008; Tison et al. 2010).

For each layer and species a local continuity equation follows and is amenable to discretization then explicit, semi- or implicit integration. If the choice is a forward Euler,

$$d\mathbb{RT}_{i,bgc}/dt = -\Delta\mathbb{F}_{i,bgc}/h^{layer} + (\mathbb{P} - \mathbb{L})_{i,bgc} \; ; \; \mathbb{F}_{i,bgc} = -v^{lam}\Delta RT^{lam}_{i,bgc} \tag{A37}$$

$$\left(\mathbb{RT}^{n+1}_{i,bgc} - \mathbb{RT}^{n}_{i,bgc}\right)/\Delta t = d\mathbb{RT}^{n}_{i,bgc}/dt \tag{A38}$$

Here *layer* signifies the reaction box, *lam* indicates the dividing barrier and concentrations on either side (e.g. *o2b*), while *n* is the current time. The computations are performed working from time averaged CICE output for the main physical drivers, along with either over-wintering initializations. We experimented with time steps ranging from 0.1 to 10 days. Light penetration is first mapped downward through the column assuming Beer's Law and single scattering with snow, ice and preexisting algae all contributing (Lavoie et al. 2005; Elliott et al. 2012). The $L^{total}$ are then constructed, carbon growth and nutrient uptake estimated and finally organisms and organics are distributed through the multiple numerical layers.

**Appendix B: Parameters**

A biogeochemical baseline simulation is defined here through the parameter list of Table B1, serving as the foundation for sensitivity tests. We adopt a strategy of unifying pack ice autotrophy across the hemispheres, based on strong commonalities from existing polar models. Settings are then diverged slightly in order to simulate fundamental taxonomic differences. This approach is very much like one which has been used to achieve high fidelity in multi-element simulation of pelagic biogeochemistry (Sarmiento et al. 1993; Moore et al. 2002 and 2004; Gregg et al. 2003).

Since ice algal populations originate in the sea, we adopt a familiar open water tracer combination to deal with internal competition. It has been applied in identical configurations to both Southern Ocean and Arctic food webs (Walsh et al. 2001 and 2004). A representative model organism is drawn from each of three major high latitude autotroph classes: diatoms, microflagellates and prymnesiophytes which may be thought of in turn as pennates, cryptophytes plus dino- and other microflagellates, and finally *Phaeocystis*. This combination is dominant in vertical sections from many sample cores (Gradinger, 1999; Fritsen et al. 2001; Lizotte, 2001). All the classes are notable producers of organics (Van Rijssel et al. 2000; Krembs et al. 2002; Schoemann et al. 2005; Underwood et al. 2010). We will focus our arguments on these few algae at the expense, for example, of centric diatoms or the coccolithophores since alternate species may not be available or else are incapable of entering the pack (Lizotte, 2001 and 2003; Walsh et al. 2004). The selected organisms are seeded into the model ice field at equal densities. Subsequent behaviors are then dictated by photo- and nutrient physiological settings according to the parameters in Table B1.

Marine ecodynamics models are normally constructed from a core of historical growth data. We borrow this technique as a means of establishing consistent global brine channel chemistry. Eppley (1972) originally reviewed maximum photosynthetic rates measured up to that time, for the entire surface sea. He was able to establish exponential temperature dependence from an envelope of data, arriving at a Q10 of about two. The study was fairly exhaustive and has not required major revision. Arrigo and Sullivan (1992) examined the photophysiology of Antarctic ice algae in culture and established pack ecosystem values working downward in temperature to account for advancing conditions in the brine. The effects of increased salinity on rate maxima were also reported and are superimposed. Additionally a functional form was proposed for photoadaptation and this too is included in our set. Arrigo et al. (1993) reduce maximum saturation intensities as a function of attenuated PAR to mimic acclimation.

This overall physiological approach was extended to the full ice column for generic Antarctic situations by both Arrigo et al. (1997) and Fritsen et al. (1998). It has also found its way into several Arctic pack ice biology models, including those in CICE (Lavoie et al. 2005; Jin et al. 2006; Deal et al. 2011; Elliott et al. 2012; Hunke et al. 2015). Meanwhile, regional polar ecologists have developed open water ecodynamics unified across the hemispheres (Walsh et al. 2001 and 2004). Similar photosynthetic and limitation parameters have simulated nutrient cycling along both the Palmer Peninsula and coastal Beaufort/Chukchi Seas. Not only are the algal seed organisms in these computations appropriate to occupy brine niches, a single model provides for geocycling in the marginal/seasonal ice domain in both hemispheres. We adopt this suite of ecosystems as a coherent conceptual foundation. Common growth, photophysiological and nutrient restriction settings apply through the total global ice column, allowing a carbon cycle-based study of the macromolecules.

The major organism classes diatoms, microflagellates and *Phaeocystis* map cleanly from open water simulations, and so will transfer conveniently back to our ocean model during coupled calculations. None are attempted here, but it is worth noting that there is a close correspondence with taxonomies from global models of intermediate complexity. For example, Moore and colleagues originally divided pelagic photosynthesizers into diatoms and smaller organisms, then segregated *Phaeocystis* (Moore et al. 2004, Wang and Moore, 2011; Wang et al. 2014). Following ice domain mechanisms (Walsh et al. 2001 and 2004), our three algal types are given similar photophysiologies, but empirically they are often dominant respectively in the ice bottom, interior and upper layers (Gradinger, 1999; Tison et al. 2010). Such ecosystem structural differences may eventually play strongly into the epontic organic chemistry. We therefore begin to discriminate algal types within the matrix in a stepwise manner.

Due to their relative size, ice diatoms are not subject to grazing (Walsh et al. 2001) and further, they are able to use the skeletal layer as a refuge (Lizotte, 2001; Lavoie et al. 2005). By contrast our smaller species experience at least some consumption pressure, and in fact it may be necessary to explain distribution features (Gradinger, 1999; Walsh et al. 2001; Schoemann et al. 2005; Tison et al. 2010). Background grazing proportion is thus increased within the bottom layer as a first order means of achieving diatom dominance. Dynamic zooplankton treatments will eventually be needed, and higher trophic levels can be accommodated simply by transferring algorithms based on high latitude water column models (Walsh et al. 2001 and 2004; Tedesco et al. 2012). Diatoms are excluded entirely from the interior here, owing to size effects and their ability to maintain vertical position (Lavoie et al. 2005) In more detailed simulations, organism dimension and pore diameters may be compared.

Although limitation terms are intentionally similar for all nondiatoms, the flagellates are more often reported from interior core sections with *Phaeocystis* above (contrast for example Fritsen et al. 2001 with Tison et al. 2010). In the present simulations no individual group is given a population dynamic head start. The Walsh approach offers potential flagellate-to-*Phaeocystis* ratios which are skewed toward the former and could be used for initialization (2004). In the current work however, early algal concentrations are set to a single value. *Phaeocystis* is included because its life history is specialized -it is known for the release of organosulfur compounds, polysaccharides and amino sugars (Van Rijssel et al. 2000; Solomon et al. 2003; Stefels et al. 2007). Polymer chemistry allows this unique organism to become effectively multicellular by forming colonies. Buoyancy may be regulated, and rising motions are sometimes observed (Schoemann et al. 2005). In future simulations we will simultaneously investigate flagellate population offsets alongside *Phaeocystis* behaviors, and competition is expected. High concentrations of the prymnesiophyte are often observed at upper levels and can likely be captured with a proper emphasis on niche adaptation.

In global open waters, dissolved organic carbon is thought to emanate from autotrophic cells primarily during disruption (Parsons et al. 1984; Carlson et al. 2000; Benner, 2002; Moore et al. 2004). Exudation is generally considered to be a secondary mode. In high latitude pelagic simulations(e.g. Walsh et al. 2004), direct emission factors are in agreement with the typical assumptions, because they are very low (order percent of carbon flow). But interesting exceptions can be cited. The diatoms are thought to release carbohydrates for aggregation (Passow et al. 1994) and sugar polymers play into ice surfactant and structural interactions (Krembs et al. 2002 and 2011; Underwood et al. 2010). *Phaeocystis* produces similar substances to form the gel matrix necessary for colony formation (Van Rijssel et al. 2000). In the Southern Ocean exudates function as siderophores and ligands (Hassler and Schoemann, 2009; Hassler et al. 2011). For present purposes, we fix macromolecular fractions at planetary average values for the biosynthesis of marine proteins, polysaccharides and lipids (Parsons et al. 1984; Wakeham et al. 1997; Elliott et al. 2014), but the background is supplemented through a set of light regulated exudation terms (Underwood et al. 2010). The aim is to reproduce and extend results from those field studies resolving ice organic material into its chemical constituents (Amon et al. 2001; Herborg et al. 2001; Thomas et al. 2001; Underwood et al. 2013).

Taken together, our parameter settings are designed to provide baseline ice biogeochemistry adequate to either the regional or global scale, in both the biogeographic and vertical dimensions (Arrigo et al. 1997; Deal et al. 2011). We introduce a minimum of taxonomic resolution (Gradinger, 1999; Fritsen et al. 2001), build local ice ecosystems and simultaneously track organic matter in the brine (Thomas and Papadimitriou, 2003). Selected baseline values are summarized in Table B1, with columns organized by algal type. Relative to our equations the collective flagellates now represent all smaller taxa. The total biological resolution is therefore three classes. Data are offered in roughly the order of appearance in the equation list. We deal with primary production as regulated by light fields then nutrient uptake, simple fractional routings and finally the production and degradation of organic macromolecules. This sequence serves as an organizing principle throughout the work.

The parameter table has been constructed based on examination of diverse individual references/source materials. Uncertainties are large; error bars of a factor of three or more in either direction accompany some of the choices. The situation is typical for modern ecodynamics simulators (Sarmiento et al. 1993; Moore et al. 2004; Lavoie et al. 2005; Elliott et al. 2012; Tedesco et al. 2012). Formal quantification lies beyond the scope of the present work, but is warranted. Major sensitivity tests described in the main text revolve around physical thicknesses, transfer velocities and light saturation dependence of organic release. For the moment emphasis has been placed upon fresh organics, but we underscore an entire network of fractional routings that must yield to dynamic channels. They will be explored later in full scale CICE modeling.

**Table B1**. Settings used to establish a baseline simulation of ice biogeochemistry, with emphasis on production of organics in the matrix. Blank cells carry information downward from their neighbor directly above. Abbreviations: NA –not applicable, Chl –chlorophyll, PAR –photosynthetically available radiation, none –dimensionless in terms of units, poly – polysaccharide. Reference abbreviations: A79 (Ackley et al. 1979), A93 (Arrigo et al. 1993), A97 (Arrigo et al. 1997), A03 (Arrigo, 2003), AS92 (Arrigo and Sullivan,1992), B02 (Benner, 2002), E72 (Eppley, 1972), E01 (Eslinger et al. 2001), EI01 (Eslinger and Iverson, 2001), E12 (Elliott et al. 2012), E14 (Elliott et al. 2014), F98 (Fritsen et al. 1998), F00 (Fung et al. 2000), J06 (Jin et al. 2006), KW95 (Kirst and Wienke, 1995), L05 (Lavoie et al. 2005), M04 (Moore et al. 2004), S97 (Smith et al. 1997), S05 (Schoemann et al. 2005), Sa05 (Sarthou et al. 2005), T95 (Thomas et al. 1995), TA06 (Tagliabue and Arrigo, 2006), W01 (Walsh et al. 2001), W04 (Walsh et al. 2004), WM11 (Wang and Moore, 2011). Comments: The low reference intensities $I_s$ are each adjusted moving upward through the column toward higher available photon fluxes according to the function provided in Arrigo et al. (1993). Photoacclimation is thus assumed to be identical among species. The growth constant is reduced as a function of salinity per the data in Arrigo and Sullivan (1992) consistent with the form offered in a follow up work (Arrigo et al. 1993). Finally, the reader may note that from $f^{spill}$ onward phyto-class distinctions disappear. Some table columns could thus be merged from this point. They are maintained for clarity of presentation.

| Quantity | Diatoms | Flagellates | *Phaeocystis* | Units | References |
|---|---|---|---|---|---|
| $R^{C/N}$ | 7 | 7 | 7 | mole/mole | W01, W04. S05 |
| $R^{C/Si}$ | 5 | NA | NA | mole/mole | A93, L05 |
| $R^{C/Fe}$ | $2 \times 10^5$ | $2 \times 10^5$ | $5 \times 10^5$ | mole/mole | F00, S05, TA06 |
| $R^{C/Chl}$ | 40 | 40 | 100 | mass/mass | A79, W01, S05 |
| $a_{Chl}$ | 0.03 | 0.01 | 0.05 | 1/m(mg/m$^3$) | AS92, W01, L05 |
| $I^s$ (low PAR) | 1.5 | 1.5 | 1.5 | W/m$^2$ | A93, A97, F98, W01 |
| $I^s$ (high PAR) | 4 | 4 | 4 | | |
| $I^s$ (pelagic) | 50 | 30 | 20 | | W01, WM11 |
| $I^{in}$ | 100 | 100 | 100 | | AS92, KW95, A03, J06 |
| $K^{NO3-}$ | 1 | 1 | 1 | μM | A93, W01, W04, S05 |
| $K^{NH4+}$ | 0.3 | 0.3 | 0.3 | | |
| $K^{Si(OH4)}$ | 3 | NA | NA | | E01, Sa05, L05 |
| $K^{Fe}$ | 100 | 100 | 10 | pM | S05, TA06 |
| $g^{pre}$ | 0.85 | 0.85 | 0.85 | 1/d | E72, AS92, W01, L05 |
| $g^{exp}$ | 0.06 | 0.06 | 0.06 | 1/C° | AS92, F98 |
| $f^{sal}$ (50 ppt) | 1 | 1 | 1 | none | AS92, A93, F98 |
| $f^{sal}$ (100 ppt) | 0 | 0 | 0 | | |
| $f^{graze}$ (bottom) | 0 | 0.9 | 0.9 | | See text |
| $f^{graze}$ (other) | NA | 0.1 | 0.1 | | A93, L05, E12, A03 |
| $f^{resp}$ | 0.05 | 0.05 | 0.05 | | A93, W01, L05 |
| $m^{pre}$ | 0.02 | 0.02 | 0.02 | 1/d | E01, EI01, J06, E12 |
| $m^{exp}$ | 0.03 | 0.03 | 0.03 | 1/C° | |
| $f^{spill}$ | 0.5 | 0.5 | 0.5 | none | M04, E12 |
| $f^{assim}$ | 0.5 | 0.5 | 0.5 | | 1-$f^{spill}$ |
| $f^{excr}$ | 0.5 | 0.5 | 0.5 | | E12 |
| $f^{remin}$ | 1 | 1 | 1 | | M04, J06, E12 |
| $k_{nit}$ | 0.015 | 0.015 | 0.015 | 1/d | J06, E12 |
| $f^{biomass}(protein)$ | 0.6 | 0.6 | 0.6 | none | P84, B02, E14 |
| $f^{biomass}(poly)$ | 0.2 | 0.2 | 0.2 | | |
| $f^{biomass}(lipid)$ | 0.2 | 0.2 | 0.2 | | |
| $k^{exude}$ | 0 | 0 | 0 | 1/d | W01, B02, M02 |
| $k^{bac}$ | 1 | 1 | 1 | 1/month | S97, T95 |